\documentclass[preprint,12pt,authoryear]{elsarticle}
\usepackage{amsmath,amsfonts,amssymb}

\newcommand{\N}{\mathbb{N}}
\newcommand{\R}{\mathbb{R}}
\newcommand{\train}{tr}

\usepackage{amssymb}
\usepackage{lineno}
\usepackage[colorlinks=true, linkcolor=blue, citecolor=blue, urlcolor=blue]{hyperref}
\usepackage{url}
\usepackage{float}
\usepackage{subcaption}
\usepackage{graphicx}
\usepackage{caption}
\usepackage[font=small]{caption}
\usepackage{booktabs}
\usepackage{siunitx}

\usepackage{glossaries}

\glsdisablehyper
\newacronym{pinn}{PINN}{Physics-Informed Neural Network}
\newacronym{piml}{PIML}{Physics-Informed Machine Learning}
\newacronym{sir}{SIR}{Susceptible-Infectious-Recovered}
\newacronym{pde}{PDE}{Partial Differential Equation}
\newacronym{ode}{ODE}{Ordinary Differential Equation}
\newacronym{mcmc}{MCMC}{Markov Chain Monte Carlo}
\newacronym{ili}{ILI}{Influenza-Like-Illness}
\newacronym{wis}{WIS}{Weighted Interval Score}
\newacronym{ae}{AE}{Absolute Error}
\newacronym{mae}{MAE}{Mean Absolute Error}
\newacronym{mse}{MSE}{Mean Squared Error}

\newcommand{\rev}[1]{{\color{black}#1}}

\journal{Epidemics}

\begin{document}

\begin{frontmatter}



\title{Forecasting Seasonal Influenza Epidemics with Physics-Informed Neural Networks}


\author[1,2]{Martina Rama}
\author[3]{Gabriele Santin}
\author[4]{Giulia Cencetti}
\author[5]{Michele Tizzoni}
\author[2]{Bruno Lepri}
\affiliation[1]{organization={Department of Information Engineering and Computer Science, University of Trento},city={Trento},state={Italy}}
\affiliation[2]{organization={Mobile and Social Computing (MobS) Lab, Fondazione Bruno Kessler (FBK)},city={Trento},state={Italy}}
\affiliation[3]{organization={Department of Environmental Sciences, Informatics and Statistics, University of Venice},city={Venice},state={Italy}}
\affiliation[4]{organization={Centre de Physique Th\unexpanded{\'e}orique, CNRS},city={Marseille},state={France}}
\affiliation[5]{organization={Department of Sociology and Social Research, University of Trento}, city={Trento},state={Italy}}





\begin{abstract}
Accurate epidemic forecasting is critical for informing public health decisions and timely interventions. While Physics-Informed Neural Networks have shown promise in various scientific domains, their potential application to real-time epidemic forecasting remains underexplored.
Here, we present SIR-INN, a hybrid forecasting framework that integrates the mechanistic structure of the classical Susceptible-Infectious-Recovered (SIR) model into a neural network architecture.
Trained once on synthetic epidemic scenarios, the model is able to generalize across epidemic conditions without retraining. From limited and noisy observations, SIR-INN infers key transmission parameters via Markov chain Monte Carlo, generating probabilistic short- and long-term forecasts.
We validate SIR-INN using national influenza data from the Italian National Institute of Health in the 2023-2024 and 2024-2025 seasons.
The model performs competitively with current state-of-the-art approaches, particularly in terms of Weighted Interval Score.
It shows accurate predictive performance in nearly all phases of the outbreak, with improved accuracy observed for the 2024–2025 influenza season. Credible uncertainty intervals are consistently maintained, while coverage metrics highlight room for improvement in uncertainty calibration.
SIR-INN offers a computationally efficient, transparent, and generalizable solution for epidemic forecasting, appropriately leveraging the framework's hybrid design. Its ability to provide real-time predictions of epidemic dynamics, together with uncertainty quantification, makes it a promising tool for real-world epidemic forecasting.
\end{abstract}




\end{frontmatter}




\section{Introduction}

In recent years, numerical modeling for epidemic forecasting has become a key tool for public health. 
Short-term forecasts of the epidemic burden, such as the expected number of cases, hospitalizations, or deaths, can inform policymaking and provide situational awareness to respond effectively with targeted intervention strategies \citep{desai2019real,lauer2021infectious}.
Over the years, different epidemic forecasting approaches have been adopted to predict the temporal dynamics of dengue \citep{johansson2019open}, Ebola \citep{viboud2018rapidd}, seasonal influenza \citep{biggerstaff2016results,brownstein2017combining}, COVID-19 \citep{sherratt2023predictive,wolffram2023collaborative} and many other diseases \citep{delvalle2018summary,holcomb2023evaluation}.

On the one hand, recent developments in the field have highlighted the numerous advantages of using ensemble modeling strategies based on large collaborative efforts \citep{reich2019accuracy, fox2024optimizing, fiandrino2025collaborative}. 
At the same time, the most effective standalone approaches generally combine statistical or machine learning techniques with conceptual models that incorporate infectious disease transmission dynamics, either explicitly through compartmental models or implicitly through factors such as seasonality or recent trends in observed data \citep{lopez2024challenges,ray2025flusion}.
However, the use of such hybrid models is still limited in real-world applications, as demonstrated by a recent review of COVID-19 modeling studies in the United States, which showed that only 13\% of the research teams adopted a hybrid methodology for pandemic forecasting \citep{nixon2022evaluation}. 

In general, hybrid methods have been developed to preserve the knowledge of the spreading mechanism, derived from mechanistic modelization, while simultaneously leveraging the data-driven learning capabilities and flexibility of machine learning models \citep{karniadakis2021physics, o2022semi, ye2025integrating}. 
Among hybrid models, a promising framework is emerging rapidly within the field of epidemiology: the \gls{pinn} \citep{raissi2019physics}.
\glspl{pinn} are capable of incorporating the observed data and mathematical models described by \glspl{pde} through a regularization mechanism endowed with the neural network.
Furthermore, compared with purely data-driven approaches, a \gls{pinn} can take advantage of few and noisy data while still ensuring robustness and generalization capabilities for its physically consistent predictions \citep{raissi2019physics,karniadakis2021physics,cuomo2022scientific}.

For these reasons, \glspl{pinn} are successfully used for a wide variety of applications such as aerodynamics, fluid mechanics, biology, and epidemiology \citep{lagergren2020biologically,mao2020physics,cai2021physics}.
Specifically, within the epidemiological context, the integration of prior epidemiological knowledge (derived from mechanistic models) enables the neural network to serve as an efficient surrogate, accurately learning the dynamics of disease spread and inferring disease-related parameters \citep{shaier2021data,bertaglia2022asymptotic,ning2023physics,ning2023epi,qian2025physics}.
However, despite the fact that the literature on epidemiological \glspl{pinn} is becoming vast and ubiquitous, the prediction of real-time epidemic scenarios with \glspl{pinn} has not been extensively explored. Some recent efforts have begun to address this gap.
\cite{madden2024deep} has proposed integrating a time series \gls{sir} model (TSIR) with \glspl{pinn} that improves both forecast and parameter inferences of measles dynamics.
\cite{berkhahn2022physics} implemented a \gls{pinn} for COVID-19 modelization and future outbreak scenario generation.
\cite{millevoi2024physics} implemented a multiple \glspl{pinn} framework investigating joint and split approaches to estimate temporal changes in model parameters and state variables, also providing a short-term forecast application on Italian COVID-19 data.
The Epidemiologically-Informed Neural Networks (EINNs) framework, introduced by \cite{rodriguez2023einns}, combines a time module and a feature module to leverage multiple data sources for both short-term and long-term forecasting. 
Finally, \cite{kharazmi2021identifiability} investigated several epidemiological models adopting the \glspl{pinn} approach that uses multiple neural networks, identifying time-dependent parameters and forecasting with uncertainty quantification.

Most of these works adopt a framework that simultaneously addresses both forward and inverse problems, fully exploiting the potential of \glspl{pinn}. Alternatively or jointly, some of the aforementioned approaches rely on extensive use of different neural networks, employing them independently to solve specific tasks (e.g., learning the compartments of the \glspl{pde} model, finding the unknown parameters) and subsequently combining their results.
In either case, when aiming for real-time forecasting, a weakness emerges: the designed framework requires neural network retraining whenever new observations (or new epidemic scenarios) become available, thus increasing both the computational cost and the delay in receiving results. 
This also restricts the generalizability of a single pre-trained neural network. Moreover, by design, the neural network evaluation produces one single deterministic output as forecast trajectories. 
As a drawback, in almost all of these cases, there is no uncertainty analysis with respect to the forecasting process, which, instead, is provided by statistical inferential approaches and stochastic simulation studies \citep{bracher2021evaluating,sherratt2023predictive}.

To address these limitations while preserving the known benefits and potential of a hybrid approach, we propose here a novel framework for epidemic forecasting, namely Susceptible-Infectious-Recovered--Informed Neural Network (SIR-INN).
In particular, we construct a single \gls{pinn} that takes advantage of the simplicity and modeling ability in describing different epidemic dynamics of the \gls{sir} model.
The prior epidemiological knowledge embedded in the neural network architecture is directly derived from the differential equations underlying the \gls{sir} compartmental model, with constant transition rates. 
Our \gls{pinn} is trained once on synthetic observations of epidemic scenarios that are close to the real ones in a temporal-epidemic domain that considers both the epidemic parameters and time as variables. This enables our model to efficiently learn a wide range of disease dynamics, generalizing its understanding to the transmission patterns of other infectious diseases. Thus, starting from a limited and noisy set of observations, the pre-trained SIR-INN is able to estimate, via the \gls{mcmc} method, the parameters that characterize the \gls{sir}-based epidemic dynamics. 
Finally, by exploiting these estimated parameters, our model can efficiently forecast future time windows. In particular, our SIR-INN solution provides (i) a single neural network trained on synthetic data, eliminating the need for retraining or relying on multiple networks and significantly limiting the computational cost of the method, while preserving the mechanistic structure of epidemic spreading; (ii) an efficient inference model for epidemiological data, combining a deterministic neural network as the underlying model with \gls{mcmc}; (iii) explicit uncertainty quantification of the estimated epidemiological parameters, finally ensuring probabilistic forecasting.

To validate our proposed framework in a real-world epidemic scenario, we use national data from the Italian seasonal influenza surveillance system, provided by the Italian National Institute of Health (ISS)\citep{iss}. 
In particular, we perform forecasting simulations four weeks and ten weeks ahead for both the 2023-2024 and 2024-2025 influenza seasons. 
Furthermore, we compare our results with state-of-the-art approaches for seasonal influenza forecasting, presented in the Influcast Hub \citep{Influcast}, and adopt the same evaluation metrics as those used by \cite{fiandrino2025collaborative}.
Such results and comparisons suggest that SIR-INN offers an efficient, cost-effective, and accurate hybrid alternative for real-world epidemic forecasting. It maintains high predictive accuracy even over long-term horizons and across multiple epidemic scenarios, while ensuring both generalizability and reliable uncertainty quantification.

The remainder of the paper is structured as follows. 
In Section \ref{sec:method}, we present the structure of the SIR-INN methodology, highlighting the three phases of model approximation, parameter inference, and forecasting. The details of the implementation of all components of the methodology are reported in Section \ref{sec:details}. The numerical results of the application of the proposed framework in seasonal influenza scenarios are provided in Section \ref{sec:results}. Finally, we summarize our work and derive our conclusions in Section \ref{sec:discussion}.

\section{Methods}
\label{sec:method}
The SIR-INN framework presented in this work mainly consists of three consecutive and related steps: the \gls{pinn} training, the inference of the disease parameters, and, finally, the forecasting of the future disease states.

In the first phase, described in detail in Section \ref{sec: SIR-PINN}, we implement a SIR-based \gls{pinn}. Specifically, we insert the ordinary differential equations of a \gls{sir} epidemic model as a regularization term into the loss function of a neural network. Then, we train the \gls{pinn} on synthetic data close to real disease scenarios, as explained in detail in Section \ref{sec:train_details}.
The following steps leverage this pre-trained \gls{pinn} with the aim of forecasting the disease spreading process starting from a few noisy observations.
In particular, Section \ref{sec: inference} (and Section \ref{sec: detail_params}) describes how, from a selected time window of observations, we infer the parameters that characterize the disease-spreading dynamics via the \glspl{mcmc} method.
Then, by evaluating our pre-trained \gls{pinn} model on these estimated parameters and on a future time window, we can perform both short-term and long-term disease forecasting. 
This last step is described in detail in Section \ref{sec: forecast}.

\subsection{\gls{sir} model approximation via \gls{pinn}}\label{sec: SIR-PINN}

\glspl{pinn} are universal function approximators efficiently endowed with physical knowledge described in terms of systems of either \glspl{ode} or \glspl{pde} \citep{raissi2019physics}.

Here, we focus on a \gls{pinn} that incorporates an epidemic spreading dynamic as physical knowledge, while fitting some synthetic data that describe plausible epidemic scenarios (see also Section \ref{sec:train_details} for details). 
In this setting, a well-known and widely used compartmental model is the \gls{sir} epidemic model \citep{kermack1927contribution}, defined by
\begin{equation}
\begin{cases}
\frac{dS(t)}{dt} &= - \beta S(t) I(t),  \\
\frac{dI(t)}{dt} &= \beta S(t) I(t) - \gamma I(t),  \\
\frac{dR(t)}{dt} &= \gamma I(t),
\end{cases}
\label{eq: model_SIR}
\end{equation}
where $S: [t_0,T] \to [0,1]$ represents the proportion of susceptibles, $I: [t_0,T] \to [0,1]$ the proportion of infected individuals, and $R: [t_0,T] \to [0,1]$ the proportion of removed individuals, with respect to the total population size $N$. Individuals are transferred between compartments through two transition rates: $\beta\in\R_{\geq 0}$, the epidemic contact or transmission rate, and $\gamma\in\R_{\geq 0}$, the removal rate (recovered or mortality). 
A fundamental threshold quantity for the analysis of disease spread is the so-called basic reproduction number $R_{0}=\beta/\gamma$, which represents the expected number of new infections produced by a single infected individual during their infectious period when introduced into a population where all subjects are susceptible \citep{heffernan2005perspectives}. 
The effective reproduction number,  $R_t$, is a time-dependent quantity that measures the number of new infections caused by an infected individual at any point in time of the outbreak. It can be estimated based on the basic reproduction number and on the proportion of susceptible in the population as $R_t= R_0 \cdot S(t)$, by assuming constant $\beta$ and $\gamma$ \citep{nishiura2009effective}.

Given the initial conditions $S(0), I(0), R(0)$ the system is fully specified. Summing the equations in \eqref{eq: model_SIR}, we obtain $\frac{d}{dt}(S(t)+I(t)+R(t))=0$, which implies that the total population is constant over time. Since the model is expressed in normalized form, it follows that $S(t)+I(t)+R(t)=1 \;\;\forall t \in [t_0,T]$.
Note also that the epidemic dynamics do not allow any effects of births and deaths on the populations since the time scale of the epidemic is assumed to be shorter with respect to the vital dynamics mechanism.

\begin{figure}[tp]
\centering
\hspace*{-0.1\linewidth}
\includegraphics[width=1.15\linewidth]{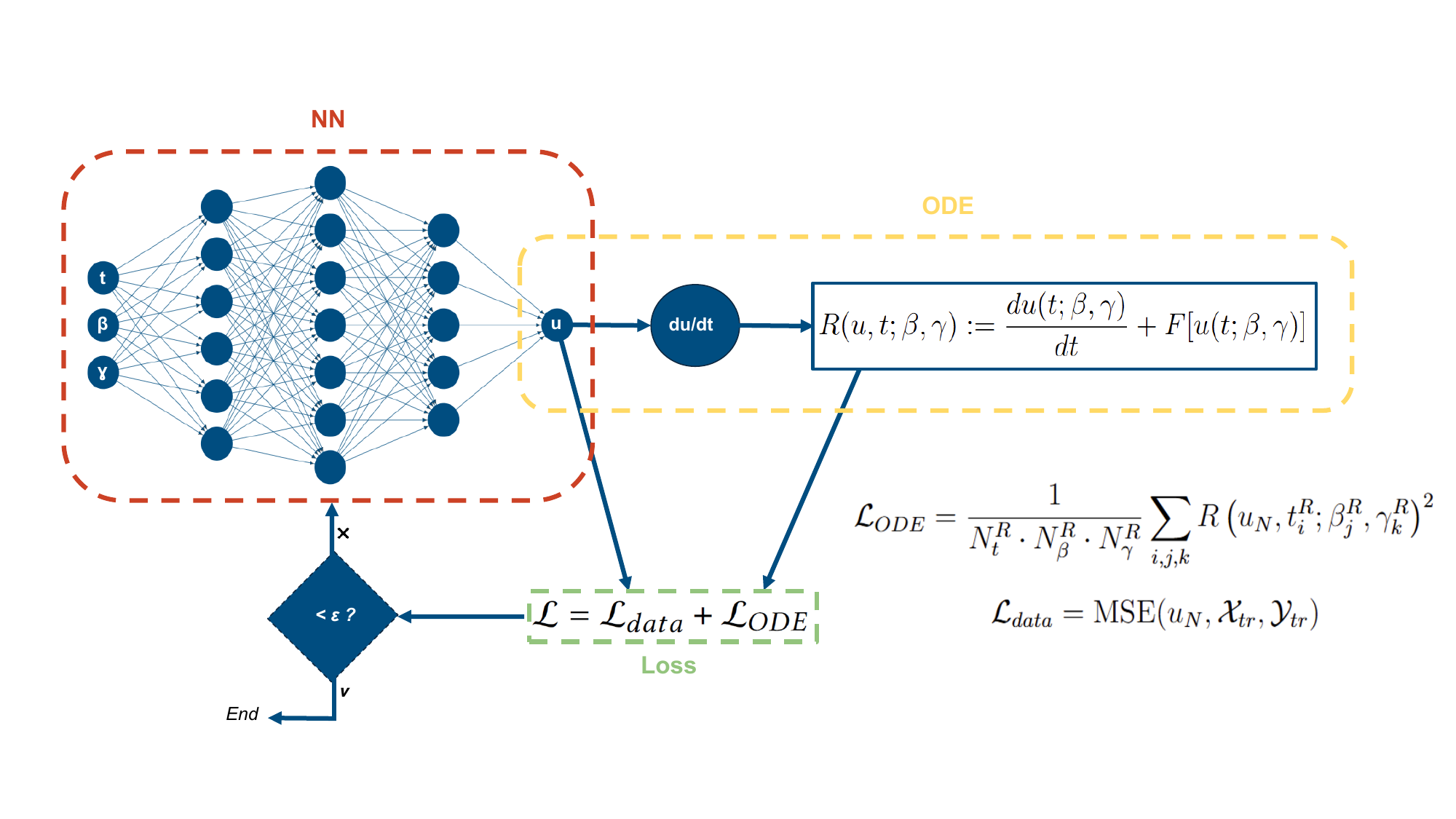}
\caption{\textbf{Diagram of the \gls{pinn} architecture for the \gls{sir} model learning.} The red-dashed frame represents the neural network which takes as inputs the time $t \in \mathcal{T}$, the disease transmission rate, $\beta\in\mathcal{B}$, the recovery rate, $\gamma\in\Gamma$, and outputs $u\in[0,1]^3$, i.e., the three normalized components of the \gls{sir} epidemic model: susceptibles, infectious, and recovered. Furthermore, the neural network is integrated with the physical knowledge of the \gls{sir} epidemic model, represented by the yellow-dashed frame, through the insertion of the \glspl{ode} system into the loss function of the neural network, becoming a physics-informed neural network. \gls{pinn} parameters are obtained by minimizing the total loss function, represented by the green-dashed frame, on the training data and satisfying a \gls{sir}-based dynamics as well.}
\label{fig:PINN_schema}
\end{figure}

The first step, illustrated in Figure~\ref{fig:PINN_schema}, is to implement a \gls{pinn} that learns the \gls{sir} model defined by Equation \eqref{eq: model_SIR}.
This \gls{ode} system can be formulated as a parametrized non-linear \glspl{ode} system
\begin{equation}
\frac{du(t;\lambda)}{dt} + F[u(t;\lambda);\lambda] = 0, \hspace{0.5cm} t \in [t_0,T],
\label{eq: ODE_general}
\end{equation}
where $\lambda:=(\beta, \gamma)\in\R_{>0}^2$ are the parameters of \eqref{eq: model_SIR}, $F[u;\lambda]$ is the parametric non-linear differential operator, and $u: [t_0,T] \times \R^2 \to [0,1]^3$, $u(t; \lambda):=(S(t), I(t), R(t))\in[0,1]^3$ is the solution map~\citep{Folland1995,Evans2022}.

With the framework defined by Equation~\eqref{eq: ODE_general}, the objective of the \gls{pinn} approach is as follows.
Given fixed model parameters $\lambda \in \mathbb{R}^m$, find a neural network $u_{N}(t;\lambda)$ that approximates the solution $u(t;\lambda)$ of the \glspl{ode} system.
We recall that a neural network $u_{N}$ is a parametric function depending on a vector $\theta$ of parameters learnable via an optimization procedure, which will be explained in the following, that finds the values of $\theta$ which steer the network output towards some desired values. Since the specific inner working of these parameters is not the focus of our paper, we omit the dependence on $\theta$ in our notation, and we refer to~\citet{lecun2015deep} and to~\citet{goodfellow2016deep} for further details.
To ensure that the neural network that approximates the solution $u(t; \lambda)$ efficiently learns a variety of epidemic scenarios, we extend the temporal domain of the function approximator to a temporal-epidemic domain that considers the epidemic parameters as variables in addition to the time variable.
Hence, we construct the \gls{pinn} as a function $u_{N}: \mathcal{T} \times \mathcal{B} \times \Gamma \rightarrow  [0,1]^3$, namely $u_N(t,\beta,\gamma)$, with the aim of approximate the solution $u(t;\beta,\gamma)$. 
Although $u_N$ is in principle defined for more general inputs, we aim at an accurate model for values in $\mathcal{T} \times \mathcal{B} \times \Gamma$, with $\mathcal{T} = [t_{min},t_{max}]$, $\mathcal{B} = [\beta_{min},\beta_{max}]$, and $\Gamma = [\gamma_{min},\gamma_{max}]$, where $0\leq t_{min}<t_{max}$, $0\leq \beta_{min}<\beta_{max}$, $0\leq \gamma_{min}<\gamma_{max}$ are suitable values that are specified in the following.

The methodology setting is data-driven, i.e., we suppose we have knowledge of a number of samples of the unknown solution, namely
\begin{align}
\label{eq: pinn_training_set}
\mathcal X_{\train}&=\{(t_i^u;\beta_j^u,\gamma_k^u), \; i=1, \dots, N_t,j=1, \dots, N_{\beta},k=1, \dots, N_{\gamma}\}\subset \mathcal{T} \times \mathcal{B} \times \Gamma,\nonumber\\
\mathcal Y_{\train}&=\{y_i=u(t_i^u; \beta_j^u,\gamma_k^u), \hspace{5pt} i=1, \dots, N_t,j=1, \dots, N_{\beta},k=1, \dots, N_{\gamma}\}, 
\end{align}
where $N_t$,$N_{\beta}$ and $N_{\gamma}$ are the total number of sampled parameters $t_i^u$,$\beta_j^u$ and $\gamma_k^u$, respectively. Note that these parameters are chosen in the same intervals $\mathcal{T}, \mathcal{B}, \Gamma$ defined before. The values $y_i$ are collected by numerical solutions of the system~\eqref{eq: ODE_general} for the corresponding parameters.

The neural network's parameters are then optimized to minimize a certain loss function, which is a measure of the network's accuracy.
In particular, \glspl{pinn} use a model-dependent, customized loss that integrates the governing equations of the \glspl{ode} system~\eqref{eq: ODE_general} while fitting the data~\citep{raissi2019physics}. More precisely, the loss has two components
\begin{equation}
\mathcal{L} = \mathcal{L}_{data} + \mathcal{L}_{ODE},
\label{eq: Loss_PINN_general}
\end{equation}
where
\begin{equation}
\mathcal{L}_{data} = \text{MSE}(u_{N}, \mathcal X_{\train},\mathcal Y_{\train})
\label{eq: LossDATA_PINN_general}
\end{equation}
is the standard \gls{mse} loss, i.e., the approximation error computed considering the training dataset~\eqref{eq: pinn_training_set} of $N_t \cdot N_{\beta} \cdot N_{\gamma}$ points,
while the physics-endowed loss $\mathcal{L}_{ODE}$ is the component that forces the neural network to approximate the \glspl{ode}.
This is done by defining a so-called residual function $R(u, t; \lambda)$ associated to the \glspl{ode}~\eqref{eq: ODE_general}, i.e.,
\begin{equation}
R(u, t; \lambda) := \frac{du(t; \lambda)}{dt} + F[u(t;\lambda); \lambda], \hspace{0.5cm} t \in [t_0,T], \hspace{0.5cm} \lambda \in\R^2_{>0},
\label{eq: PINN_res_general}
\end{equation}
in such a way that the physics-endowed loss penalizes the solutions with a large residual, i.e.
\begin{equation}
\mathcal{L}_{ODE} = \frac{1}{N^R_t \cdot N^R_{\beta} \cdot N^R_{\gamma}} \sum_{i,j,k} R\left(u_N, t^R_i; \beta^R_j, \gamma^R_k\right)^2,
\label{eq: LossODE_PINN_general}
\end{equation}
where the points are taken in the so-called collocation set
\begin{equation}
\label{eq: pinn_collocation_set}
\mathcal X_{coll}=\{(t^R_i; \beta^R_j, \gamma^R_k), \; i=1, \dots, N^R_t,j=1, \dots, N^R_{\beta},k=1, \dots, N^R_{\gamma}\}.
\end{equation}
The construction of this physics-informed loss function enables the addition of a regularization term that drives the network to learn the underlying model structure, simultaneously fitting the data, and preserving the initial conditions if they are specified. 
Compared to purely data-driven training, this mechanism allows us to benefit from one simple feedforward neural network, with few layers and neurons, and to train it on small amounts of data~\citep{raissi2019physics,karniadakis2021physics}.
Although the two loss terms in~\eqref{eq: Loss_PINN_general} are sometimes weighted with different coefficients in order to balance their relative importance, we found in our results that no noticeable difference can be obtained in this way (see~\ref{sec:param_losses}). We thus keep an unweighted definition for the loss in our framework.

\subsection{Parameters inference}
\label{sec: inference}

As a second step of our methodology, we employ an \gls{mcmc} \citep{neal1993probabilistic,richey2010evolution} method to extract, from a few noisy observations, the epidemic features that describe the underlined disease spreading process.

In particular, once our SIR-INN is trained on synthetic data approximating credible epidemic scenarios as described in Section \ref{sec: SIR-PINN} (and in Section \ref{sec:train_details}), we use \gls{mcmc} to estimate the epidemic parameters that drive the \gls{pinn} solution to be as close as possible to the observations.
As fundamental parameters that characterize the outbreak dynamics, we choose the initial time $\tau_0\in\R_{\geq 0}$ of the observed spreading process, that is the time at which we start to observe the data, and the two transition rates of the \gls{sir}-based model: the transmission rate and the removal rate, $\beta\in\R_{\geq 0}$ and $\gamma\in\R_{\geq 0}$, respectively. 
It is important to note that at this stage the weights of the neural network are fixed, and the trained SIR-INN represents a function $u_{N}: \mathcal{T} \times \mathcal{B} \times \Gamma \rightarrow [0,1]^3$.
Then, the \gls{mcmc} method is implemented assuming, as priors, uniform distributions for all the parameters and, as reference model, our pre-trained \gls{pinn}. Furthermore, we chose a likelihood function that reflects the nature of epidemiological data: a Poisson distribution with a function of the model output serving as its mean. For details about this step, please refer to Section \ref{sec: detail_params}.

Once we run the \gls{mcmc} algorithm for a selected time window of observations, we obtain a Markov chain of the estimated parameters. We then select $1000$ posterior samples of the tails, namely: $\hat{\tau}_0, \hat{\beta}, \hat{\gamma}$, from which we can extract summary statistics and perform probabilistic forecasting.

\subsection{Forecasting}
\label{sec: forecast}
The last step of our methodology, that is forecasting the future state of the epidemic, is the most crucial but perhaps the simplest to implement. Indeed, it only amounts at evaluating the pre-trained SIR-INN model with the parameters estimated via \gls{mcmc}, as explained in the previous section.

Namely, for each observation window we obtain $\hat{\tau}_0,\hat{\beta},\hat{\gamma}$, as in~Section \ref{sec: inference}, and evaluate the SIR-INN model on $\hat\beta, \hat \gamma$ parameters (with their uncertainty) and on the full time window, starting at time $0$.
We then translate time $0$ to the estimated initial time $\hat{\tau}_0$ to obtain the predicted $S$, $I$, and $R$ components of interest. The resulting values can be used both for short- and long-term probabilistic forecasting, just selecting a forecasting window size.

Iterating this procedure over each time rolling-window of observations, it provides forecasts over the entire time period of interest.

\section{Implementation and evaluation details}
\label{sec:details}
All simulations were coded in the Python programming language using the libraries \texttt{PyTorch} version $2.6.0$~\citep{PyTorch2019}, \texttt{Pymcmcstat} package version $1.9.1$~\citep{Miles2019}, \texttt{seaborn} library version $0.13.2$~\citep{Waskom2021}. Validation metrics were computed using the R programming language, specifically with the \texttt{scoringutils} package version $1.2.2$~\citep{bosse2022evaluating}.

\subsection{\gls{pinn} architecture and training}\label{sec:train_details}

Regarding the neural network architecture, we leverage the regularization mechanism of the \glspl{pinn} to benefit from simple feedforward neural networks, minimizing as much as possible the computational cost derived from the usual deep neural networks \citep{raissi2019physics,karniadakis2021physics}. 
We implemented a single \gls{pinn} consisting of only three hidden layers, and an overall architecture of $[3, 16, 32, 16, 3]$, where each value denotes the number of neurons per layer, from input to output.
As activation functions, we use smooth differentiable functions commonly used for \glspl{pinn} \citep{cuomo2022scientific}: hyperbolic tangent for the hidden layers and a sigmoid activation function for the output layer, to have outputs in the range $[0,1]$. This data normalization is commonly adopted for training neural networks: in our case, this requires normalizing the data by the total population size $N$, as $(S/N,I/N, R/N)$ components, during training. This normalization is consistent both with the \gls{ode} system \eqref{eq: model_SIR} embedded in the neural network via loss function \eqref{eq: LossODE_PINN_general}, and with the second step of our framework (i.e., parameter inference), where the model outputs are compared with incidence data, which are inherently expressed as proportions with respect to the population size.

As described in Section \ref{sec: SIR-PINN}, to train our SIR-INN, we chose synthetic data for the training set and the collocation set. In particular, since we generated them to construct the training set \eqref{eq: pinn_training_set}, we used the same samples for the collocation set \eqref{eq: pinn_collocation_set}. Hence, following the same notation of Section \ref{sec: SIR-PINN}, we considered $\mathcal X_{tr} \equiv \mathcal X_{coll}$.
For the time component, $t_i^u$, we used weekly time data in a time window of $600$ days; from $t_1^u = 0$ to $t_{86}^u = 595$. 
Regarding the ranges of parameters, we define a uniform grid of size $50$ for the removal rate $\gamma \in [1/12,1/2.5]$, and the transmission rate $\beta \in [0.12, 0.45]$. Then we restricted the values with linear constraints around the diagonal of the grid (that is, around $R_0 = \beta/\gamma = 1$) obtaining $1068$ pairs of values $(\beta_j^u,\gamma_k^u), \hspace{5pt} j=1, \dots, N_{\beta},k=1, \dots, N_{\gamma}$, where $N_{\beta} \cdot N_{\gamma} = 1068$. These intervals were selected to represent plausible epidemiological conditions for seasonal influenza \citep{carrat2008time,biggerstaff2014estimates}. 
In particular, the corresponding infectious period $1/\gamma$ ranges from approximately $2$ to $12$ days, while the basic reproduction number $R_0=\beta/\gamma$ spans the interval $(0.75,2.5)$. These values were chosen with the aim of covering a sufficiently broad domain that could both plausibly contain parameter values inferred from real influenza surveillance data and maintain the neural network generalization capabilities. Indeed, the training domain contains both subcritical regimes (i.e., $R_0<1$) in which epidemics do not occur, and regimes where $R_0 > 1$ characterized by epidemic growth that allows the framework to deal with different influenza seasons.
A visualization of all the parameter pairs included in the training set is provided in \ref{sec: B_ood} (Figure \ref{fig:trainvsinfer}).

To construct the $\mathcal Y_{tr}$ training set~\eqref{eq: pinn_training_set}, for each pair of parameters, $( \beta_j^u,\gamma_k^u)$, we first generated the \gls{sir} components via \textit{odeint}. 
This is done considering a daily time window of $600$ days for $t_i^u$, a population size $N = 10^6$, and the \gls{sir} initial conditions $(S(0),I(0),R(0)) = (N-1,1,0)$. Second, we normalized all the values by the total population size, which correspond to obtain the solutions $u(t_i^u; \beta_j^u,\gamma_k^u)$, defined in Section \ref{sec: SIR-PINN}.
Third, we restricted the generated samples to the weekly time window, coherently with the $\mathcal X_{tr}$ training set~\eqref{eq: pinn_training_set}. We train our \gls{pinn} for $7500$ epochs using the Adamax optimization algorithm, with a learning rate of $0.001$. The neural network is trained with the same batches of $100$ training, validation and collocation points, with train size = $0.9$.

All of these parameters and synthetic data choices are chosen according to plausible scenarios of rapidly transmitted respiratory infection spreading, while trying to generalize them to better train our neural network, increasing its generalization capabilities. 
For more details about the approximation capabilities of the network, including an out-of-distribution analysis, and the suitability of the training set showing that the epidemiological parameters inferred from real influenza data lie well within the training domain, we refer the reader to~\ref{sec: B_ood}.
Regarding the choice of $N$, it is intentional and does not affect the general results of the model, since the SIR dynamics is scale-invariant for sufficiently large populations. Moreover, we consider $N$ as an effective interacting population, since assuming the full Italian population to be susceptible and well mixed would be unrealistic (see also~\ref{sec:population_size} for a sensitivity analysis on $N$). Similarly, although the \gls{pinn} is trained on deterministic synthetic data, with stochasticity incorporated at the parameter inference stage rather than at training time, we verify in~\ref{sec:stochastic_training} that introducing stochastic perturbations during training results in no significant improvement across all evaluation metrics, confirming that the deterministic training strategy is sufficient. Furthermore, although population conservation $S(t)+I(t)+R(t)=1$ is not explicitly enforced during training, we verify in~\ref{sec:N_const} that the trained \gls{pinn} satisfies it within a \gls{mae} of $10^{-4}$, confirming that the network implicitly learns to respect this physical constraint.

\subsection{Inference of disease parameters}
\label{sec: detail_params}

Parameter estimation is obtained via \gls{mcmc}, as described in Section \ref{sec: inference}.
The reference data are \gls{ili} incidence observations, which are calculated as weekly reported cases per $1000$ patients. We rescale it to our reference population of $N=10^6$ (i.e., we multiply the measured incidence by $N/1000$).

As the network $u_N(t, \beta, \gamma)$ outputs normalized $S,I,R$ values, we first define a function $G$ extracting the corresponding estimated weekly incidence.
We fix a time grid $\tau_j=\tau_0+7j$, $j=0, 1, \dots$ depending on an unknown alignment time $\tau_0\in[0,600]$, and define
\begin{equation}\label{eq:weekly_aggregation}
G(j;\tau_0, \beta, \gamma)
= \left(u_N^{(0)}(\tau_{j-1},\beta,\gamma) - u_N^{(0)}(\tau_{j},\beta,\gamma)\right) \cdot N,
\end{equation}
where the superscript $u_N^{(0)}$ denotes the first output component, i.e., the estimated value of $S$.

Parameter estimation now amounts at estimating $\tau_0, \beta,\gamma$ in $G$ from the measured incidence data.
For this, we consider a rolling time window of $M=5$ weeks of observations, let $y_1, \dots, y_5$ be the rescaled incidence per $N$ people measured in these weeks, and assume
\begin{equation}\label{eq:poisson_likelihood}
y_j \sim \text{Poisson}(\mu_j),
\quad
\mu_j = G(j;\tau_0, \beta, \gamma),
\end{equation}
i.e., a Poisson distribution which reflects the counting nature of the epidemiological data, with mean modelled by the \gls{pinn}-estimated incidence.

We note that if $\tau_0$ was kept fixed over the entire estimation period it would correspond to an estimated elapsed time from the epidemic start, but this would assume an exact, fixed-parameter \gls{sir} to exactly reproduce the whole sequence of data. Since this is an unrealistic assumption, and since $\tau_0$ is re-estimated over each $M=5$ time-window, it rather plays the role of a time-alignment factor.

The \gls{mcmc} algorithm now samples from the posterior distribution
\begin{equation}
p(\tau_0, \beta, \gamma \mid \mathbf{y}) \propto p(\mathbf{y} \mid \tau_0, \beta, \gamma) \cdot p(\tau_0, \beta, \gamma),
\label{eq:posterior}
\end{equation}
where $\mathbf{y} = (y_1, \ldots, y_5)$ are the observations, $p(\mathbf{y} \mid \tau_0, \beta, \gamma)$ is the Poisson likelihood, and $p(\tau_0, \beta, \gamma)$ is the prior distribution.
The corresponding log-likelihood is
\begin{equation}
\log p(\mathbf{y} \mid \tau_0, \beta, \gamma) = \sum_{j=1}^{M} \left[ y_j \log \mu_j - \mu_j - \log \Gamma(y_j + 1) \right],
\label{eq:log_likelihood}
\end{equation}
where $\Gamma(\cdot)$ is the gamma function.
As mentioned in Section \ref{sec: inference}, we use uniform and static priors for all parameters, i.e.
\begin{equation}
\beta \sim \mathcal{U}(0.12, 0.45), \quad \gamma \sim \mathcal{U}(1/12, 1/2.5), \quad \tau_0 \sim \mathcal{U}(0, 400),
\label{eq:priors}
\end{equation}
where the bounds are chosen to encompass biologically plausible ranges for influenza transmission, in line with the training set choice (Section \ref{sec:train_details}). This choice reflects the rolling-window nature of our inference framework: parameters are independently estimated at each forecast round without assuming temporal continuity, and we lack reliable prior information about the parameter distributions for each weekly window.
In ~\ref{sec:Gaussian priors}, we additionally report an analysis using weakly informative Gaussian priors that lead to quantitatively similar results.

For each observation window (i.e., round), the \gls{mcmc} sampling is performed for $10,000$ iterations with initial parameter values set to $(\tau_0, \beta, \gamma)|_{\text{init}} = (200, 0.28, 0.24)$, namely, the midpoints of the intervals in \eqref{eq:priors}. For each round we obtain a posterior from which we discard the first $9000$ iterations as burn-in. We obtain $1000$ posterior samples (i.e., $\hat{\tau}_0, \hat{\beta}, \hat{\gamma}$) from which we extract summary statistics and do probabilistic forecasting (Section~\ref{sec: forecast}) by finally converting SIR-INN outputs - in terms of estimated $S$, $I$, and $R$ components - to incidence values, coherently with the data.

We further remark that the choice of $10,000$ iterations is robust: extending to $50,000$ iterations does not substantially affect the results (\ref{sec:chain_length}).

All these steps are implemented using the Python package \texttt{Pymcmcstat}, version $1.9.1$~\citep{Miles2019}, that can run \gls{mcmc} on any model implemented by a Python function, which in our case is simply the pre-trained physics-informed neural network implemented in \texttt{PyTorch} (version: $2.6.0$).

\subsection{Epidemiological data}\label{sec:epi_data}
We used seasonal influenza national data provided by the Italian National Institute of Health (ISS)\citep{iss} through the RespiVirNet surveillance system. 
These consist of weekly Influenza-Like-Illness (ILI) incidence observations reported by sentinel doctors at both national and regional levels (with the exception of the Valle d'Aosta and Calabria regions for the season 2023-2024), starting from the season 2003-2004 to the most recent, 2024-2025. 
In particular, we perform simulations for both the seasons 2023-2024 and 2024-2025, with influenza surveillance running from November to May, starting from week $42$ in a given year and continuing until week $17$ of the following year.
All data is publicly available on GitHub at: \cite{github_RespiVirNet}.

\subsection{Forecasting evaluation metrics}
\label{sec:metrics}
We evaluate our SIR-INN framework forecasting performances using several metrics, employed and described in detail in \cite{fiandrino2025collaborative}. Further details are provided in~\ref{sec:error_metrics_defs}.

\paragraph{\gls{ae}} It is computed as the absolute value of the difference between the median forecast and the actual corresponding value.
The \gls{mae} is the average of the AE over different time steps. We also implemented it by averaging for all the forecasting time windows.

\paragraph{\gls{wis}} It is an approximation of the Continuous Ranked Probability Score (CRPS) \citep{bracher2021evaluating} that evaluates both the accuracy of the forecast median solution and of the prediction intervals in containing the actual observations. The metric is defined for a given prediction interval of a model forecast, generalizing the interval score to multiple prediction intervals.
We refer to~\ref{sec:error_metrics_defs} for the exact choice of the corresponding parameters.

\paragraph{Coverage} It measures the calibration of a model and is defined as the fraction of times a prediction interval contains the actual data.
Notice that in a well-calibrated model, the coverage values should closely match the nominal levels of the predictive intervals. 
For example, observations will be included exactly $90\%$ of the time in the prediction interval $90\%$. We also computed the coverage for both the $50\%$ and $90\%$ prediction intervals of our forecasts.

\section{Results}
\label{sec:results}
As discussed in Section~\ref{sec:epi_data}, to validate our methodology we focus on disease forecasting using national data from the Italian influenza surveillance.
In particular, to increase both the complexity and the relevance of the task, we put ourselves under the same conditions as the teams that participated in the Influcast Hub challenge 2023-2024 \citep{Influcast}. 
Specifically, we perform four weeks ahead forecasting simulations, as outlined in Section \ref{sec:4weeks}, and we employ the same validation metrics adopted in \cite{fiandrino2025collaborative} to ensure a fair comparison between our model and all the models of the participating teams. 
For the corresponding quantitative results, we refer to Section \ref{sec:PINNvsInflucast}. 
Moreover, in Section \ref{sec: res_params_inf} we show the results on parameter inference obtained by our model using a methodology \gls{mcmc}. 
In addition, in Section \ref{sec:10weeks} we analyze the performance of our approach in a long-term forecasting task.
All experiments were carried out using data from two different influenza seasons (years 2023-2024 and 2024-2025), to illustrate the ability of our SIR-INN hybrid methodology to generalize across different epidemic scenarios without retraining the neural network. 

\subsection{Parameters inference results}
\label{sec: res_params_inf}
As explained in Section \ref{sec: detail_params}, after we have selected an ideal size for the observation time window, $M = 5$, we perform an \gls{mcmc} simulation (i.e., round) every time we observe new data, in a rolling-window manner.
For example, at week $51$, we use five data points from the weeks $47$-$51$ to estimate the parameters that best align the \gls{sir}-based \gls{pinn} output to the observed \gls{ili} incidence in this window. 
In line with the Influcast Hub, we performed simulations starting with four weeks of observation (also for the 2023-2024 influenza season) and considering retrospective weekly updates of past observations, as the system provides new surveillance data. 
In the end, the total number of simulations is equal to the total number of observations (i.e., the total number of weeks in a season) minus four, which is 24 for both seasons. 
Note that for the first simulation we used only four observations, as they are the only available.

\begin{figure}[tp]
\centering
\hspace*{-0.1\linewidth}
\includegraphics[width=1.15\textwidth,height=10cm]{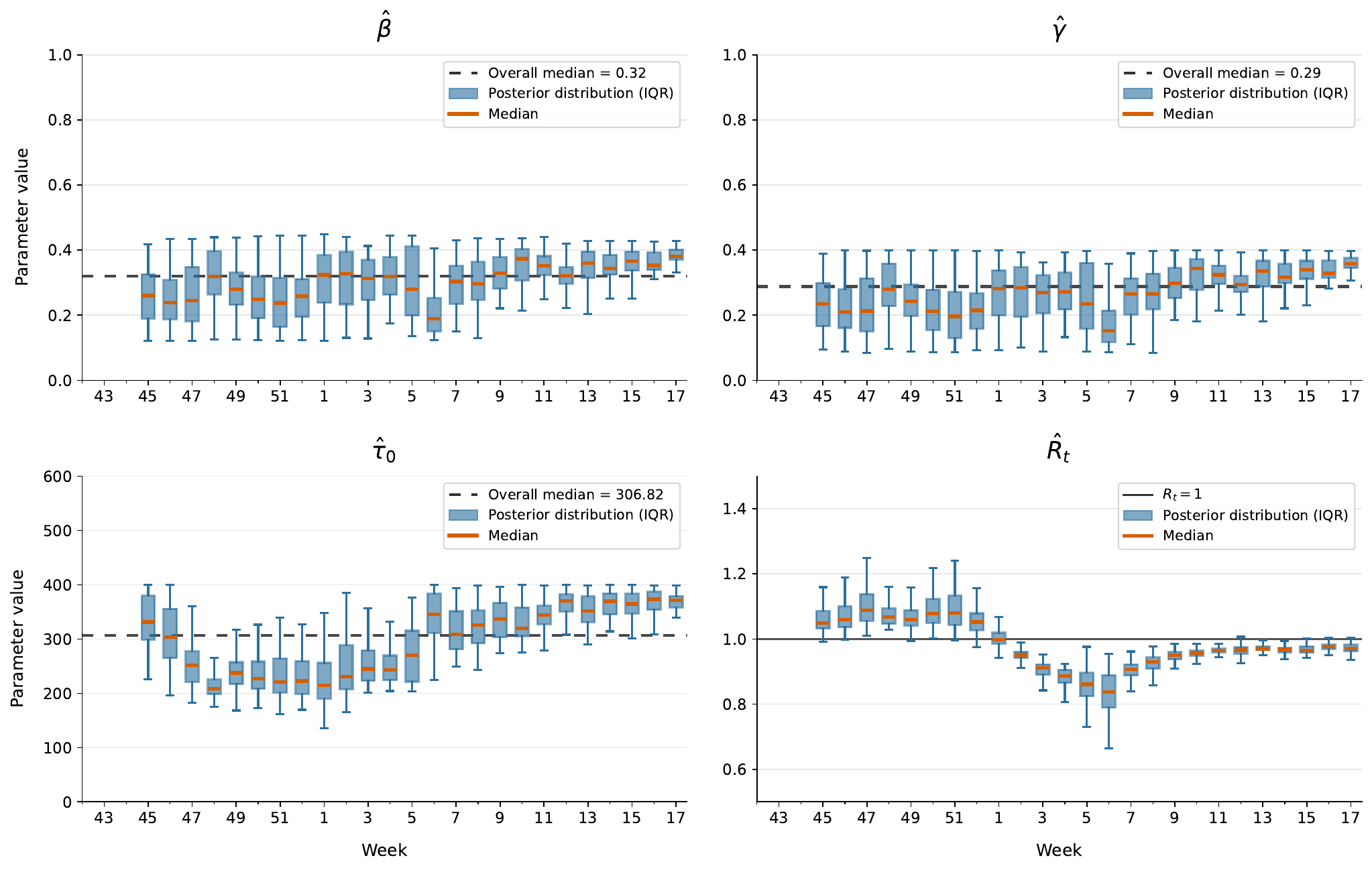}
\caption{\textbf{Behavior of the \gls{sir}-based model parameters estimated via \gls{mcmc} - 2023-2024 Italian seasonal influenza.} Posterior distributions of the inferred parameters $\hat{\beta}$, transmission rate, $\hat{\gamma}$, recovery rate, $\hat{\tau}_0$, observation alignment time, and $\hat{R}_t$, effective reproduction number, across epidemic weeks. Each box represents the interquartile range (IQR, 25th–75th percentile), with whiskers extending to the 5th and 95th percentiles; the orange horizontal line within each box denotes the weekly posterior median. The dashed line in the panels for $\hat{\beta},\hat{\gamma},\hat{\tau}_0$ indicates the overall posterior median across all weeks. For $\hat{R}_t$, the solid horizontal line marks the threshold $R_t = 1$, separating epidemic growth ($R_t>1$) from decline ($R_t<1$).}
\label{fig:Influcast_params_inference}
\end{figure}

\begin{figure}[tp]
\centering
\hspace*{-0.1\linewidth}
\includegraphics[width=1.15\textwidth,height=10cm]{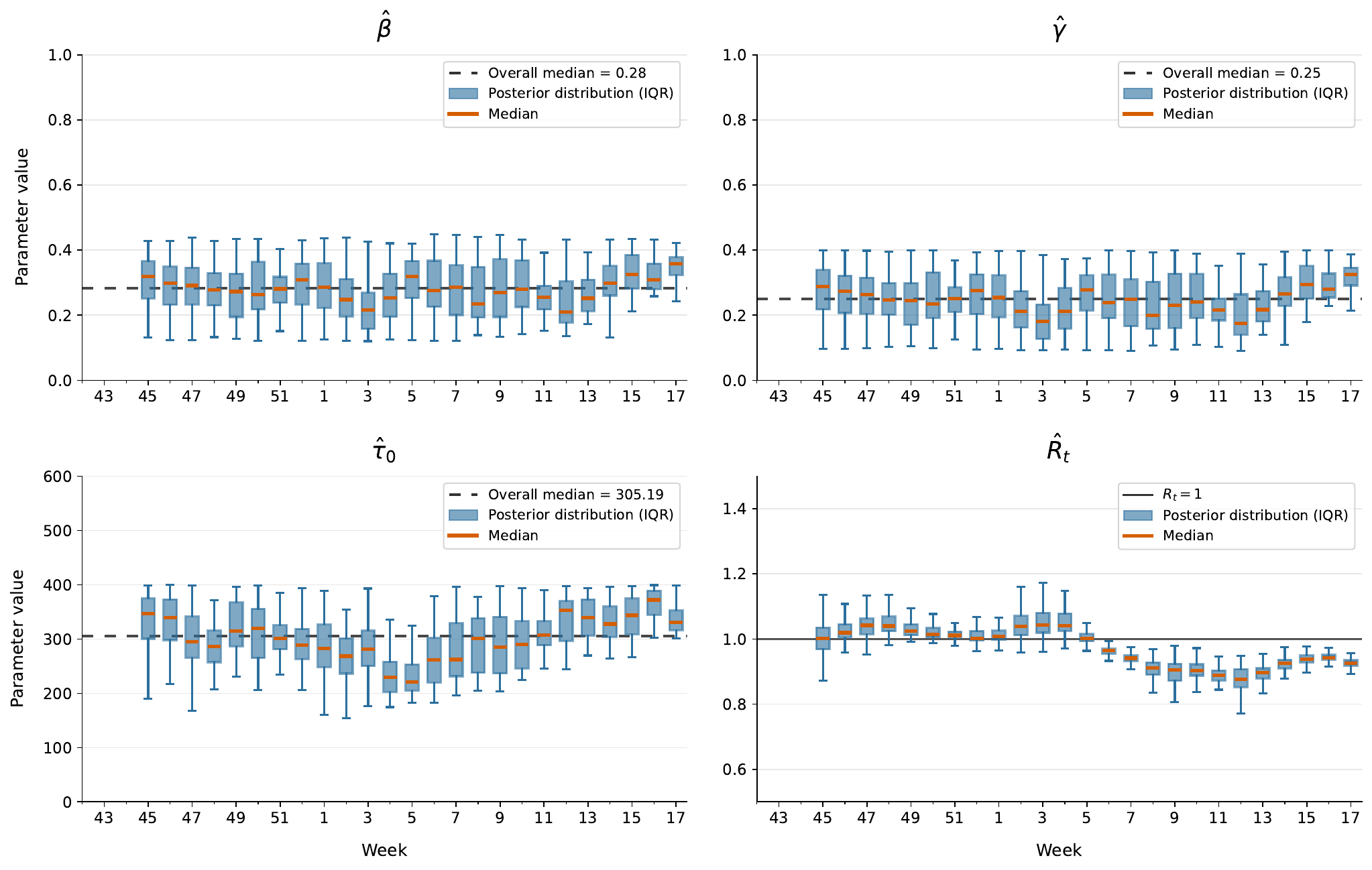}
\caption{\textbf{Behavior of the \gls{sir}-based model parameters estimated via \gls{mcmc} - 2024-2025 Italian seasonal influenza.} Posterior distributions of the inferred parameters $\hat{\beta}$, transmission rate, $\hat{\gamma}$, recovery rate, $\hat{\tau}_0$, observation alignment time, and $\hat{R}_t$, effective reproduction number, across epidemic weeks. Each box represents the interquartile range (IQR, 25th–75th percentile), with whiskers extending to the 5th and 95th percentiles; the orange horizontal line within each box denotes the weekly posterior median. The dashed line in the panels for $\hat{\beta},\hat{\gamma},\hat{\tau}_0$ indicates the overall posterior median across all weeks. For $\hat{R}_t$, the solid horizontal line marks the threshold $R_t = 1$, separating epidemic growth ($R_t>1$) from decline ($R_t<1$).}
\label{fig:Influcast_params_inference_25}
\end{figure}

Figure \ref{fig:Influcast_params_inference} and Figure \ref{fig:Influcast_params_inference_25} show the posterior distributions (specifically, the $1000$ posterior samples of the chain tails) of the estimated parameters $\hat{\beta},\hat{\gamma},\hat{\tau}_0$ and the posterior distributions of $\hat{R}_t$ for all simulations (that is, weeks), for the season 2023-2024 and 2024-2025, respectively.
From the plots, we observe that the overall median values of the  parameters are similar across the two seasons, and their values remained stable as the seasons progressed. 
The median disease transmission rates, $\beta$, are close to $0.30$ and the median removal rates, $\gamma$, are (slightly lower) close to $0.27$ day$^{-1}$, 
leading to an average $R_0$ equal to $1.13$, for the 2023-2024 influenza season, and $1.14$, for the 2024-2025 (see Figure~\ref{fig:Influcast_params_inference_all_R0} and the relative section for additional information). 
Regarding the observation alignment time, $\hat{\tau}_0$, its median throughout the weeks is close to day $300$, both for the 2023-2024 and 2024-2025 seasons.

For each epidemic week, fixing a posterior sample, the effective reproduction number $\hat{R}_t$, coherently with the definition given in Section \ref{sec: SIR-PINN}, is computed as:
\begin{equation}
    \hat{R}_t = \hat{R}_0 \cdot u_N^{(0)}(\hat{\tau}_{M-1},\hat{\beta},\hat{\gamma})
\end{equation}
where $\hat{R}_0 = \hat{\beta} / \hat{\gamma}$ is the estimated basic reproduction number, and $u_N^{(0)}(\hat{\tau}_{M-1}
,\hat{\beta},\hat{\gamma})$ is the SIR-INN predicted susceptible fraction of the population at the current time $\hat{\tau}_{M-1}=\hat{\tau}_0+7\cdot(M-1)$ (see Section \ref{sec: detail_params}), depending on the observation alignment time $\hat{\tau}_0$ estimated from the corresponding posterior sample.\\
Regarding $\hat{R}_t$, the results are consistent with the expected epidemic dynamics for both seasons. While in 2023-2024 the \gls{ili} incidence peaked at weeks 51-52, in 2024-2025 the peak arrived a month later, at weeks 4 and 5. For the first season (Figure \ref{fig:Influcast_params_inference}), both the posterior median and the interquartile range (IQR) remain above the epidemic threshold $\hat{R}_t = 1$ up to week 52, indicating sustained transmission, while at week 1 the median reaches $\hat{R}_t \approx 1$ and from week 2 onwards both the median and the interquartile range fall below the threshold, suggesting that the outbreak had entered the decline phase. 
A similar pattern is observed for the 2024-2025 season (Figure \ref{fig:Influcast_params_inference_25}), although the growth phase is more prolonged: the posterior median stays above $\hat{R}_t = 1$ up to week 4, the median reaches the threshold at week 5, and from week 6 onwards both the median and the interquartile range lie below 1. However, during the early weeks of this season, the posterior distributions exhibit greater uncertainty, with the interquartile range crossing the threshold $\hat{R}_t = 1$ from below, reflecting a less definitive signal of epidemic growth compared to the 2023-2024 season. Overall, the temporal evolution of $\hat{R}_t$ across both seasons is coherent with the observed \gls{ili} incidence curves, capturing the transition from the epidemic growth phase to the declining phase with epidemiologically plausible timing.

For the same analysis performed with a chain of length $50,000$, please refer to Section~\ref{sec:chain_length}.

Even if the best parameter values are re-estimated every week, as new surveillance data are reported and previously published data are updated, our model consistently identifies a narrow range of plausible values that remain almost constant across the season. 
The largest fluctuations are observed for the estimated $\hat{\tau}_0$, which the model adjusts to capture the temporal dynamics of the outbreak, which were different in the two seasons, as we highlighted before.  

Figure \ref{fig:trainvsinfer} shows that the posterior medians of these inferred parameter pairs $(\hat \beta, \hat \gamma)$ fall well inside the training parameter area, demonstrating that our choice for the ranges of the training parameters was appropriate. 
Furthermore, the parameter values of $\hat \beta$ and $\hat \gamma$ fall in line with the typical values measured for seasonal influenza. In particular, an average infectious period of about 4 days ($ \gamma = 0.25$ day$^{-1}$), and a basic reproductive number of $R_0 \sim 1.2$ correspond to the traditional estimates of these parameters for seasonal flu \citep{carrat2008time,biggerstaff2014estimates}. The values of the estimated parameters and their stability between weeks are encouraging, even if this does not necessarily imply a global epidemiological identifiability and interpretability.
Thus, although this is not explicitly enforced, we see that the data and the structure based on \gls{sir} allow the model to effectively learn the epidemiological role of the two parameters, $\beta$ and $\gamma$. 

\subsection{Forecasting results}
From the results derived from the estimation of $\beta,\gamma$, and $\tau_0$ parameters via \gls{mcmc}, we obtain Markov chains, referred to each time window of observations (i.e., round).
Hence, we perform probabilistic forecasting in the form of quantiles with prediction intervals based on model outcomes, on each evaluation window, defined as the union of the observations time window and the forecast time window. 
In particular, for each \gls{mcmc} chain we select $1000$ tail samples to draw from the posterior, then we evaluate our model on these samples, computing the quantiles as measures of uncertainty.

\subsubsection{Four-weeks-ahead forecasting}
\label{sec:4weeks}
In a first set of experiments, we generate forecasts four weeks ahead, according to the Influcast Hub forecasting standards \citep{Influcast}. 
This means that, similarly to the parameter inference experiments, we update our forecasts each time the epidemic curve is updated with a new observation, removing the oldest one within a time window of five data points. 
We consider a fixed size of four weeks for the forecast time window.
Note that for the last four simulations, we forecast on a shorter time window since we do not have any other future observations to compare with.

\begin{figure}[tp]
\centering
\includegraphics[width=1\textwidth,height=10cm]{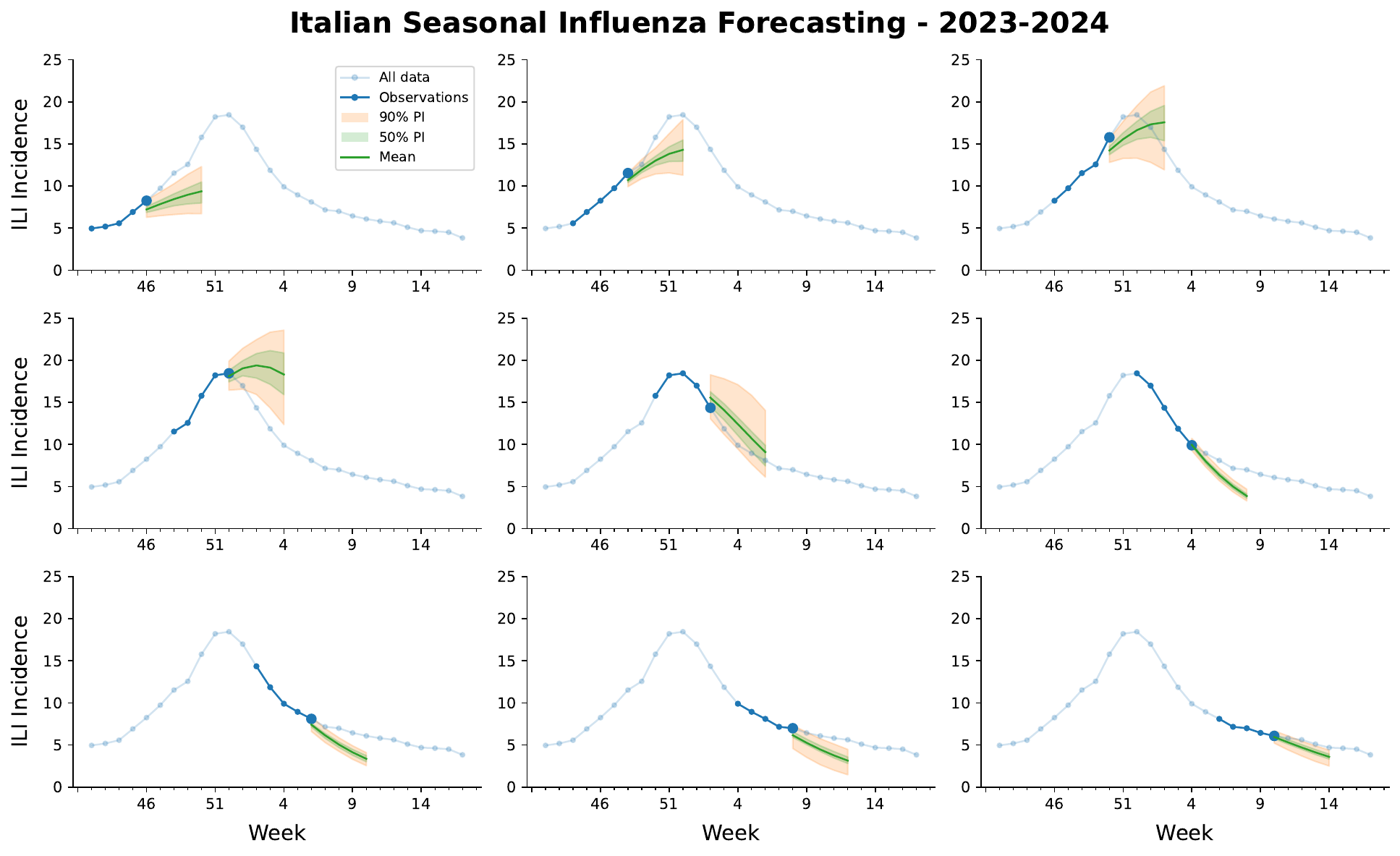}
\caption{\textbf{Seasonal influenza forecasting four weeks ahead via SIR-INN.} The ground truth data, in blue, represent the weekly observations at national level of the 2023-2024 seasonal influenza. The highlighted blue data represent the time window of observations, with fixed length of $5$, from which the \gls{sir}-based \gls{pinn} infers the epidemic parameters and then forecasts on the following four weeks ahead. The mean-model forecasts are represented with green lines and the $50\%$ and $90\%$ percentile intervals are shaded, respectively, in green and orange error bars.}
\label{fig:Influcast_forecasting}
\end{figure}

\begin{figure}[tp]
\centering
\includegraphics[width=1\textwidth,height=10cm]{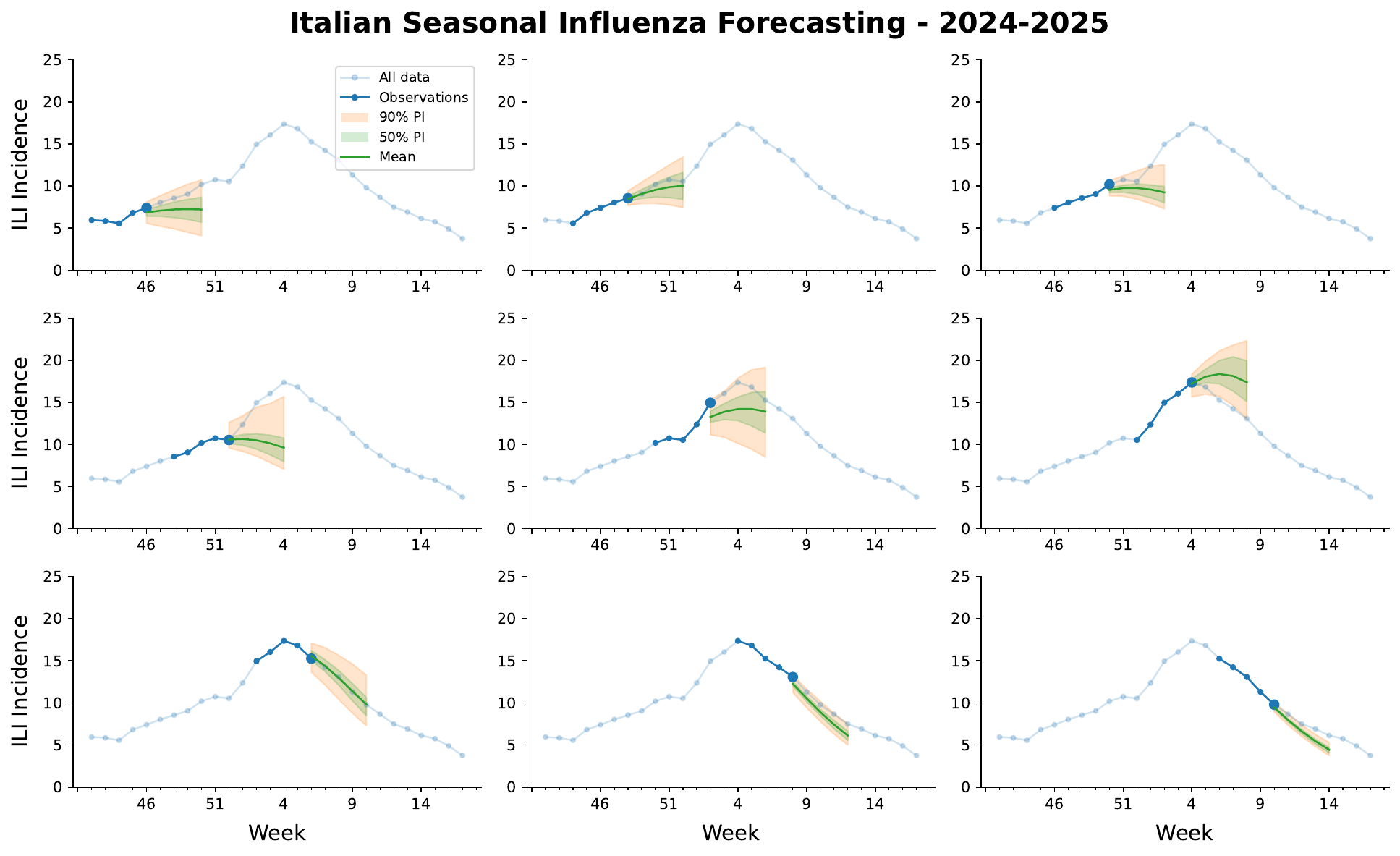}
\caption{\textbf{Seasonal influenza forecasting four weeks ahead via SIR-INN.} The ground truth data, in blue, represent the weekly observations at national level of the 2024-2025 seasonal influenza. The highlighted blue data represent the time window of observations, with fixed length of $5$, from which the \gls{sir}-based \gls{pinn} infers the epidemic parameters and then forecasts on the following four weeks ahead. The mean-model forecasts are represented with green lines and the $50\%$ and $90\%$ percentile intervals are shaded, respectively, in green and orange error bars.}
\label{fig:Influcast_forecasting_25}
\end{figure}

Figures \ref{fig:Influcast_forecasting} and \ref{fig:Influcast_forecasting_25} show the results for nine representative time windows and for the 2023-2024 and 2024-2025 seasons, respectively. 
Note that the error bars and the mean results of SIR-INN start from the last reported incidence observation since our methodology based on \gls{pinn} produces uncertainty throughout the evaluation window.
The complete results obtained considering all the time windows are shown in \ref{sec: A_fit_plots}, Figure \ref{fig:Influcast_fit_all}, for the 2023-2024 season, and in Figure \ref{fig:Influcast_fit_all_25}, for the 2024-2025 season.

In the 2023-2024 influenza season (Figure \ref{fig:Influcast_forecasting}), we can see that SIR-INN is fairly accurate in the initial phase of the outbreak, despite the inherent challenges associated with the limited number of observations available at that stage. 
However, the forecast tends to underestimate the observed incidence in this initial period, as shown by the forecasts at weeks 46 and 48. 
As the peak time approaches in weeks 51 and 52, we can see that the error bars become larger and the $90\%$ percentile interval contains the current data and almost all of the following four observations. 
In particular, at week 50 the model almost correctly identifies the occurrence of the peak, which is the most critical and challenging phase, even if the actual incidence value is slightly underestimated.
The declining epidemic phase that follows the peak is well predicted by SIR-INN, in its initial phase (week 2). 
However, as the outbreak fades, our model rapidly begins to underestimate the real observations, forecasting a swift extinction of the newly reported cases. 
The forecasting accuracy increases again in the final phase of the season, as the curve approaches the baseline values once more (weeks 8-10).

The results for the 2024-2025 season (Figure \ref{fig:Influcast_forecasting_25}) show an increased forecasting accuracy, compared to the previous year.
The early outbreak phase, the peak time, and the decreasing phase are consistently estimated with good accuracy. 
In particular, the crucial forecast of the peak time is improved with respect to the results of the 2023-2024 season.  
The peak incidence is slightly underestimated when projected from week $2$, and slightly overestimated at week $4$. 
The predicted peak time in both cases is nearly identical to the real one, and the general trend is well captured by our model.
However, at week $50$, we can see that our model is unable to accurately predict the later increase in the \gls{ili} incidence curve. Nevertheless, it seems to be able to predict the constant trend of the subsequent two weeks and with $90\%$ interval a possible increasing of the incidence function. Indeed, by week $52$, the model appears surprisingly to infer that the current peak may represent only a relative infection peak, thus anticipating a potential subsequent rapid increase in incidence. 
For the peculiarity of this first less pronounced peak of infection, weeks 49-52 are the most challenging predictions that are still in line with most of the forecasts reported on the Influcast platform \citep{Influcast}.
Finally, our model effectively captures the decreasing phase of the outbreak. 
This clearly emerges from the forecasts at weeks $6$ and $8$, where the SIR-INNN mean value, represented by the green line, almost perfectly fits the unseen observations, and the $50\%$ percentile intervals contain almost all the data. Only the latest observations are slightly underestimated by our model, as shown by the results at week $10$. 

Table \ref{tab: SIRINN 2seasons} describes the forecast performance of our model compared to the baseline, which is used as a reference model in \cite{fiandrino2025collaborative}, for both seasons. 
The results are obtained from national-level four-weeks-ahead forecasting using several metrics, aggregating forecasts from all rounds and all the horizons. The number of rounds represents the number of simulations considered for evaluation. In line with the Influcast hub challenge \citep{Influcast}, we considered a total number of $20$ rounds for the 2023-2024 season, starting the first four-weeks-ahead forecasting from the $46$th week of $2023$ and finishing with the last forecast simulation at week $13$th of $2024$. 
For the 2024-2025 season, we began to evaluate the performances of the simulation whose last observation corresponds to the $45$th week of $2024$. 
The number of horizons considered for the evaluation is four, corresponding to the size of the forecast time window.
A detailed description of the evaluation metrics is reported in Section \ref{sec:metrics}.

\begin{table}[tbp]
\centering
\resizebox{\textwidth}{!}{
\begin{tabular}{lcccccc}
\toprule
\textbf{Model} & \textbf{Influenza Season} & \textbf{N. of Rounds} & \textbf{MAE} & \textbf{WIS} & \textbf{Coverage 50\%} & \textbf{Coverage 90\%} \\
\midrule
Baseline & 2023/2024 & 20 & 2.68 & 1.90 & \textbf{0.49} & \textbf{0.66} \\
SIR-INN & 2023/2024 & 20 & \textbf{2.28} & \textbf{1.63} & 0.11 & 0.39 \\
\midrule
Baseline & 2024/2025 & 21 & 2.54 & 1.71 & 0.14 & 0.52 \\
SIR-INN & 2024/2025 & 21 & \textbf{1.98} & \textbf{1.36} & \textbf{0.24} & \textbf{0.58} \\
\bottomrule
\end{tabular}
}
\caption{\textbf{Forecasting performance of the SIR-INN model, compared to the baseline.} The results are obtained from the four-weeks-ahead forecasting at national level, across the two influenza seasons, in terms of \gls{mae} of the median forecast, \gls{wis}, $50\%$ and $90\%$ coverage. Bold values indicate the best performance for each metric within the corresponding season.}
\label{tab: SIRINN 2seasons}
\end{table}

From the results shown in Table \ref{tab: SIRINN 2seasons} we can see that, in both seasons, SIR-INN consistently outperforms the baseline in terms of point forecast accuracy, as confirmed by lower \gls{mae} and \gls{wis} values.
Specifically, the \gls{mae} of the median forecast decreases from $2.68$ to $2.28$ in 2023-2024 and from $2.54$ to $1.98$ in 2024-2025, confirming an improvement in forecast precision for the 2024–2025 season compared to 2023–2024.
Furthermore, the \gls{mae} of the median forecast, the \gls{wis} values, and the coverages of the 2024-2025 season are notably optimal with respect to the baseline as well as the previous season, meaning that the accuracy of the median forecast and the prediction intervals in bounding the actual observations is also greater compared to the baseline model.
Moreover, although the coverage values remain somewhat below the ideal nominal levels, their increase suggests that the model exhibited improved calibration by transitioning from the previous influenza season to the most recent. 

\subsubsection{Comparison with Influcast models}
\label{sec:PINNvsInflucast}
Since we performed four-week forecast simulations of Italian seasonal influenza 2023-2024 according to the requirements of the Influcast challenge \citep{Influcast}, we implemented the same validation metrics adopted in \cite{fiandrino2025collaborative} to evaluate the accuracy of our model relative to those of the other teams. 
Table \ref{tab:SIINNvsInflucast} presents the aggregated performance, for all rounds and horizons, of all models under consideration, as in \cite{fiandrino2025collaborative}, with the addition of our model's performance (last row). 
Regarding the performance evaluation of all the Influcast models, the relative values of the \gls{mae} of the median forecast and the \gls{wis} are computed by comparing pairs of models that participate in the same forecast round. These values are then normalized by dividing by the corresponding value of the baseline model. In addition, these values incorporate both national-level seasonal influenza predictions and aggregated regional forecasts, in contrast to national-level results reported in Section \ref{sec:4weeks}. 
In this way, each individual model's result accounts for both regional and national aggregated outcomes of all other models, and it is expressed relative to the baseline.
Consequently, relative values below $1$ indicate better performance than the baseline, while values above $1$ indicate worse performance. 
The ensemble model is obtained by combining all the model forecasts, and it consistently outperforms the baseline model on all results, as detailed in \cite{fiandrino2025collaborative}, which also provides further information on the single models and evaluation metrics. Regarding our model results, the relative \gls{mae} of the median forecast and the WIS are computed by comparison with the results of all other models and normalized by the corresponding baseline values. 
Note that SIR-INN's relative results are evaluated with respect to the other models, whereas the reverse does not hold, as we did not officially participate in the challenge. Consequently, the ensemble model also does not include our model's predictions. Moreover, the values related to our model are all derived from national-level predictions, rather than regional ones.

\begin{table}[tbp]
\centering
\resizebox{\textwidth}{!}{
\begin{tabular}{lcccccc}
\toprule
\textbf{Model} & \textbf{N. of Rounds} & \textbf{Relative MAE} & \textbf{Relative WIS} & \textbf{Coverage 50\%} & \textbf{Coverage 90\%} \\
\midrule
Mechanistic-1        & 19 & 0.56 (1st) & 0.54 (1st) & 0.75 & 0.89 (1st) \\
Mechanistic-2        & 20 & 1.86 & 1.64 & 0.04 & 0.49 \\
Mechanistic-3        & 20 & 0.58 (2nd) & 0.58 (3rd) & 0.47 (3rd) & 0.70 (3rd) \\
Mechanistic-4        & 19 & 0.73 & 0.73 & 0.16 & 0.37 \\
Semi-mechanistic-1   & 16 & 1.80 & 2.14 & 0.06 & 0.27 \\
Semi-mechanistic-2   & 14 & 0.84 & 1.04 & 0.19 & 0.30 \\
Statistical-1        & 20 & 0.99 & 1.09 & 0.17 & 0.43 \\
Statistical-2        & 19 & 0.77 & 0.85 & 0.11 & 0.31 \\
\midrule
Ensemble             & 20 & 0.59 (3rd) & 0.57 (2nd) & 0.51 (1st) & 0.80 (2nd) \\
Baseline             & 20 & 1.00 & 1.00 & 0.49 (2nd) & 0.66 \\
\midrule
SIR-INN              & 20 & 1.03 & 0.98	& 0.11 & 0.39 \\
\bottomrule
\end{tabular}
}
\caption{\textbf{Forecasting performance of the SIR-INN model compared to Influcast models.} The results are obtained from the four-weeks-ahead forecasting at national level in terms of relative \gls{mae} of the median forecast, relative \gls{wis}, $50\%$ and $90\%$ coverage.}
\label{tab:SIINNvsInflucast}
\end{table}

Taking into account the results of our model, outlined in the last row of Table \ref{tab:SIINNvsInflucast}, the relative \gls{mae} and \gls{wis} values are close to the baseline values ($1.03$ and $0.98$, respectively), resulting slightly better than the baseline in terms of relative \gls{wis}, and the overall performance remains comparable to that of the other models. Specifically, in terms of relative \gls{mae} and \gls{wis}, our model outperforms two other models: \textit{Mechanistic-2} (1.86, 1.64), \textit{Semi-mechanistic-1} (1.80, 2.14). Additionally, it outperforms \textit{Statistical-1} (1.09) and \textit{Semi-mechanistic-2} (1.04) in term of \gls{wis}.
Regarding the coverage, although the values are below the ideal nominal levels, our model still compares well with the others. 
It performs better than \textit{Mechanistic-2} (0.04), \textit{Semi-mechanistic-1} (0.06), and similarly to \textit{Statistical-2} (0.11), for the coverage $50\%$. Furthermore, in terms of coverage $90\%$, SIR-INN improves over \textit{Mechanistic-4} (0.37), \textit{Semi-mechanistic-1} (0.27), \textit{Semi-mechanistic-2} (0.30), and \textit{Statistical-2} (0.31).
In summary, our model shows overall performance in line with the participating Influcast models \citep{Influcast}, usually ranking above the bottom three. This suggests that its implementation could positively contribute to the prediction of future seasonal influenza epidemics when included in the ensemble forecasts.

\subsubsection{Ten-weeks-ahead forecasting}
\label{sec:10weeks}
In addition, we analyze the performances of our SIR-INN on long-term forecasting. 
The experimental setup is identical to the one implemented for the short-term forecasts, with the only difference being a different size for the forecasting time window. 
In this case, we select a range of ten weeks instead of four weeks. 

\begin{figure}[tp]
\centering
\includegraphics[width=\textwidth,keepaspectratio]{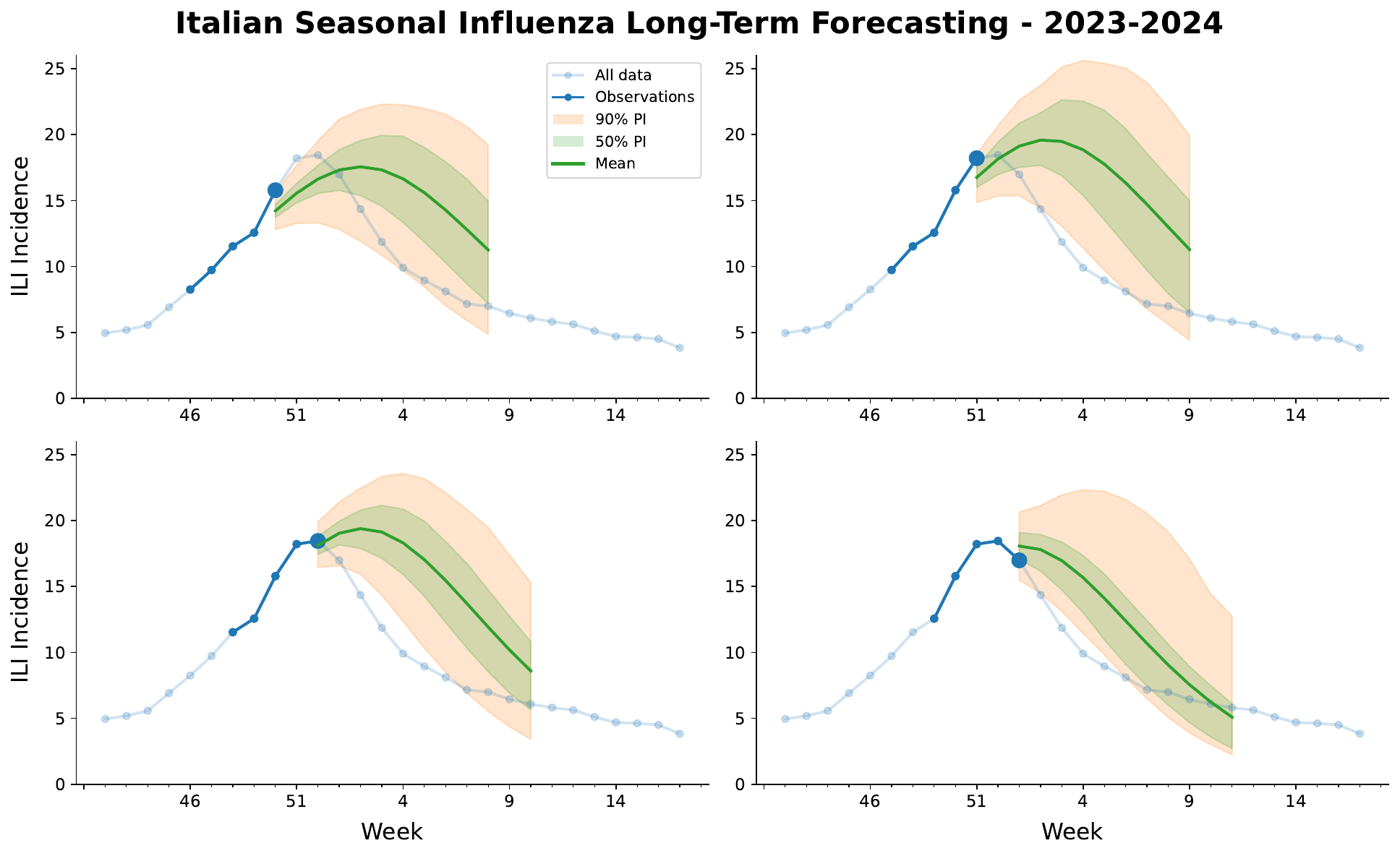}
\caption{\textbf{Seasonal influenza forecasting ten weeks ahead via SIR-INN.} The ground truth data, in blue, represent the weekly observations at national level of the 2023-2024 seasonal influenza. The highlighted blue data represent the time window of observations, with fixed length of $5$, from which the \gls{sir}-based \gls{pinn} infers the epidemic parameters and then forecasts on the following ten weeks ahead. The mean-model forecasts are represented with green lines and the $50\%$ and $90\%$ percentile intervals are shaded, respectively, in green and orange error bars.}
\label{fig:Influcast_long_forecasting_24}
\end{figure}

The results for the 2023-2024 season obtained considering four consecutive time windows of observations during the peak phase are shown in Figure \ref{fig:Influcast_long_forecasting_24}. For the results of the 2024-2025 season, see Figure \ref{fig:Influcast_long_forecasting_25}, in \ref{sec: A_fit_plots}.

The results of the 2023-2024 influenza season, shown in Figure \ref{fig:Influcast_long_forecasting_24}, illustrate how our model is able to partially capture the epidemic trend beyond the short-term horizon in this delicate phase, although without providing a fully accurate estimate. In particular, at week 50, two weeks before the peak, all the future ten unknown incidence values fall within the $90\%$ percentile interval.
In all scenarios depicted in Figure \ref{fig:Influcast_long_forecasting_24}, the forecast curve exhibits a broader shape relative to the curve derived from interpolated observations, tending to underestimate the data points before the peak and, conversely, overestimate the observations once the peak has passed. 
The peak time is consistently predicted later than the actual one, although the discrepancy remains within a maximum of four weeks. 
Notably, shortly after the epidemic peak (week 1), the model rapidly adapts to the changing dynamics, being able to provide accurate longer-term forecasts of the declining phase of the incidence curve, with relatively narrow prediction intervals.

Even in the case of long-term forecasting, the results presented in \ref{sec: A_fit_plots}, Figure \ref{fig:Influcast_long_forecasting_25}, are better for the 2024–2025 season than for the 2023–2024 one. 
The $90\%$ prediction interval contains almost all observations for the peak scenarios and, at week 5, immediately after the peak, almost all the observations fall within the narrow $50\%$ percentile interval, proving again the ability of the model to quickly adjust its predictions.
In addition, the model's ability to predict the maximum \gls{ili} incidence increases as the epidemic peak approaches.

\section{Discussion}
\label{sec:discussion}
In this work, we have introduced a novel framework for epidemic forecasting, namely SIR-INN, based on \glspl{pinn}.
Specifically, we endowed a single neural network with prior epidemiological knowledge, leveraging the well-known \gls{sir} compartmental model with constant transition rates.
The neural network has been trained only once on temporal-epidemic domains close to realistic epidemic scenarios. 
In this way, starting from a limited set of noisy observations, our \gls{pinn} is able to infer, via \gls{mcmc}, the parameters that characterize a \gls{sir}-based epidemic dynamics.
Following this inference step, we used our \gls{pinn} to forecast the dynamics of epidemics in future time windows.

We used national data from the Italian seasonal influenza surveillance, provided by the Italian National Institute of Health (ISS) \citep{iss}, as done by previous studies on influenza forecasting \citep{zhang2015social, perrotta2017using, brownstein2017combining}. 
Our results, obtained from a direct application of the framework to national data from the influenza seasons 2023-2024 and 2024-2025, confirm the good forecasting abilities of our hybrid methodology.

Despite differences in weekly observations and predictions of associated parameters, our results indicate that both seasonal influenza outbreaks showed similar epidemiological characteristics in terms of transmissibility and average infectious period. 
Such epidemiological features are shown to be underpinned by a \gls{sir} model, with disease transmission rate and removal rate values piecewise constant close to $0.3$ and an average basic reproduction number $R_0$ of approximately $1.135$. Furthermore, the temporal evolution of the effective reproduction number $R_t$ 
confirmed consistent epidemic dynamics across both seasons, with $R_t$ exceeding the threshold of $1$ during the growth phase and declining below it at epidemiologically plausible turning points, coherently with the observed \gls{ili} incidence curves.

We then performed four-week-ahead forecasting simulations based on inferred parameters, employing the same validation metrics adopted by \cite{fiandrino2025collaborative}. 
From the qualitative results, we can see that for the 2023-2024 influenza season, and especially for the 2024-2025 season, the mean forecasting solutions during the early period and the decreasing stage are often fairly accurate. This is particularly relevant considering that providing accurate forecasts during the early phase of an outbreak is challenging, primarily due to the limited number of available observations, which are often noisy. Even more challenging is the estimation of the incidence peak. Although the model does not achieve high accuracy in forecasting the season peak, it appears to anticipate its occurrence reasonably well. 
In general, Figures \ref{fig:Influcast_forecasting} and Figure \ref{fig:Influcast_forecasting_25} suggest that our model is more effective in capturing the temporal pattern of the influenza season 2024-2025. This impression is also confirmed by the results in Table \ref{tab: SIRINN 2seasons}, which reports the \gls{mae} of the median forecast, \gls{wis}, and $50\%$ and $90\%$ coverage for both seasons.
In general, our SIR-INN framework appears to outperform the baseline used as a reference model in \cite{fiandrino2025collaborative}, particularly in terms of \gls{mae} and \gls{wis} for both the seasons, and also in terms of coverages for the 2024-2025 season. This makes our model a possible competitive alternative to the baseline for seasonal influenza forecasting.
These findings are also supported by the results of the seasonal influenza forecast for 2023-2024 presented in Table \ref{tab:SIINNvsInflucast}, where all models participating in the Influcast four weeks ahead forecast challenge \citep{Influcast} were compared to ours. These results show that the performances of our model are comparable to those of current state-of-the-art models. In particular, the SIR-INN performs better in terms of the relative \gls{wis}, thereby demonstrating the accuracy of the prediction with respect not only to the median outcome but also to the prediction intervals.
In summary, these results suggest that our framework could meaningfully enhance the forecasting of successive seasonal influenza epidemics, increasing the ensemble's predictive performance.

In addition, in Section \ref{sec:10weeks}, we analyze the performance of our model in long-term forecasting, specifically ten weeks in advance. In Figure \ref{fig:Influcast_long_forecasting_24} and Figure  \ref{fig:Influcast_long_forecasting_25}, we can see that although the influenza season is in its most challenging phase---peak estimation---and only a few observations are available, our model is able to partially capture the epidemic trend. This is particularly relevant considering that long-term forecasting often fails when using fully data-driven and neural network-based approaches \citep{rodriguez2024machine,ye2025integrating}.

To further contextualize these results, we conducted an additional comparison with a classical mechanistic \gls{sir} model and a hybrid PINN-ODE pipeline, reported in detail in~\ref{sec:SIRvsODE}. The comparison shows that the proposed SIR-INN framework achieves competitive performance across all metrics, and in particular outperforms both the fully mechanistic and hybrid mechanistic versions in terms of \gls{mae} and \gls{wis}, while maintaining a significantly lower computational cost. These findings suggest that the relative advantages of fully mechanistic, fully neural, and hybrid approaches may depend on the specific forecasting context, including the epidemic phase, the forecast horizon, and the evaluation metric.

Together, these findings align with what is increasingly emphasized in the forecast epidemiology literature. Adopting a hybrid approach that integrates the mechanistic understanding of disease transmission with the learning and generalization capabilities of neural networks appears to be a promising direction to improve forecast performance \citep{rodriguez2024machine,ye2025integrating,qian2025physics}.
The developed framework allows us to benefit from both approaches, thus maximizing forecasting capabilities in both short-term and long-term predictions. 
On the one hand, to preserve the underlying dynamic structure of influenza transmission, we use the simplicity of the \gls{sir} model, supported by theoretical studies, with fixed epidemiological parameters. On the other hand, the neural network component allows us to exploit data-driven patterns, improving the flexibility and generalization ability of the model \citep{raissi2019physics,rodriguez2024machine}.
Furthermore, we take additional advantage of the hybrid nature of the framework by implementing an \gls{mcmc} method, separately from the neural network, to automatically capture the temporal variability of the \gls{sir} parameters through independent local estimation at each forecasting round. While this rolling re-estimation allows the model to rapidly adapt to the changing dynamics, it also implies that the inferred parameters should be interpreted as locally consistent approximations rather than global epidemiological constants characterizing the full season. Our results suggest that even complex epidemic dynamics may reflect a simple, shared mathematical structure that deserves consideration in purely ML-based forecasting models \citep{ye2025integrating}.

The resulting framework, despite the combination of diverse and potentially complex hybrid components, remarkably preserves simplicity, explainability, efficiency, and cost-effectiveness. 
Our SIR-INN consists of a single neural network trained only once on synthetic data, in contrast to the majority of machine learning (and hybrid) approaches for epidemic forecasting \citep{kharazmi2021identifiability, rodriguez2023einns, millevoi2024physics,qian2025physics}. 
This makes the method highly flexible and generalizable. The neural network does not require retraining each time a new observation is available or when a new influenza season occurs. 
The SIR-INN framework also allows us to obtain parameter distributions independently of the neural network training, thereby eliminating the need for retraining it while allowing for the quantification of forecast uncertainty. 
Performing Uncertainty Quantification (UQ) is essential in epidemiological forecasting, as it contributes to model robustness and supports more reliable decision-making in public health. Given the inherent variability and noise of epidemiological data, capturing forecast uncertainty is necessary to avoid overconfidence and misinterpretation of the model output \citep{funk2019assessing,bracher2021evaluating,sherratt2023predictive}. 
However, in neural network–based approaches where the output is a single trajectory, incorporating this uncertainty remains challenging and is therefore still rarely implemented in the current literature \citep{linka2022bayesian,shi2025survey}. 
Our work can be seen as an initial and promising attempt toward incorporating uncertainty into forecasting within this class of hybrid approaches. However, further work is still needed in this direction. The model recurrently exhibits overconfidence, resulting in overly tight percentile intervals and consequently underperforming in terms of coverage. This behavior is less pronounced around the peak phase, where wider predictive intervals reflect the inherent difficulty of peak estimation, ensuring greater reliability. Nevertheless, this represents a relevant shortcoming that should be taken into account when interpreting the probabilistic forecasts.

In summary, our proposed framework, namely SIR-INN, provides an efficient and accurate surrogate for learning epidemiological models, enabling less costly forecasting while maintaining both generalizability and accuracy, even in long-term predictions. Given its simple design, the model offers valuable insight into more efficient and interpretable implementations of hybrid models for forecasting other dynamical systems with uncertainty quantification. 
A natural and immediate extension of this work is its application to other \gls{sir}-based epidemic models. In fact, it would only require retraining the neural network once after selecting parameter ranges that are better aligned with the specific disease under study. 
Furthermore, an interesting direction for generalization is the prediction of epidemics coupled with human behavior changes. In such cases, the individual approaches—the mechanistic modeling and the neural network component—remain limited by their partial perspectives on the system, failing to benefit from one another during the forecasting process.
In conclusion, our work represents an initial step toward an increased integration of AI tools with mechanistic epidemic models to improve forecasting approaches and contributes to the establishment of new standards in hybrid modeling.

\section*{Code Availability}
\noindent
\rev{The code to reproduce the results of this paper is publicly available at \href{https://github.com/martina-rama/SIR-INN}{https://github.com/martina-rama/SIR-INN}.}

\section*{CRediT authorship contribution statement}
\noindent
\textbf{Martina Rama:} Conceptualization, Data curation, Formal analysis, Methodology, Software, Validation, Visualization, Writing – original draft, Writing – review and editing.
\textbf{Gabriele Santin:} Conceptualization, Data curation, Formal analysis, Methodology, Software, Supervision, Validation, Visualization, Writing – review and editing.
\textbf{Giulia Cencetti:} Conceptualization, Methodology, Supervision, Validation, Writing – review and editing.
\textbf{Michele Tizzoni:} Conceptualization, Supervision, Validation, Writing – review and editing.
\textbf{Bruno Lepri:} Conceptualization, Supervision, Validation, Writing – review and editing.


\section*{Acknowledgements}
\noindent
Bruno Lepri acknowledges the support of the PNRR ICSC National Research Centre for High Performance Computing, Big Data, and Quantum Computing (CN00000013), under the NRRP MUR program funded by NextGenerationEU.

\bibliographystyle{elsarticle-harv}
\bibliography{bib_file}

\appendix

\clearpage
\newpage
\section{Additional plots}
\label{sec: A_fit_plots}

\begin{figure}[H]
\centering
    \includegraphics[width=0.85\textwidth,keepaspectratio]{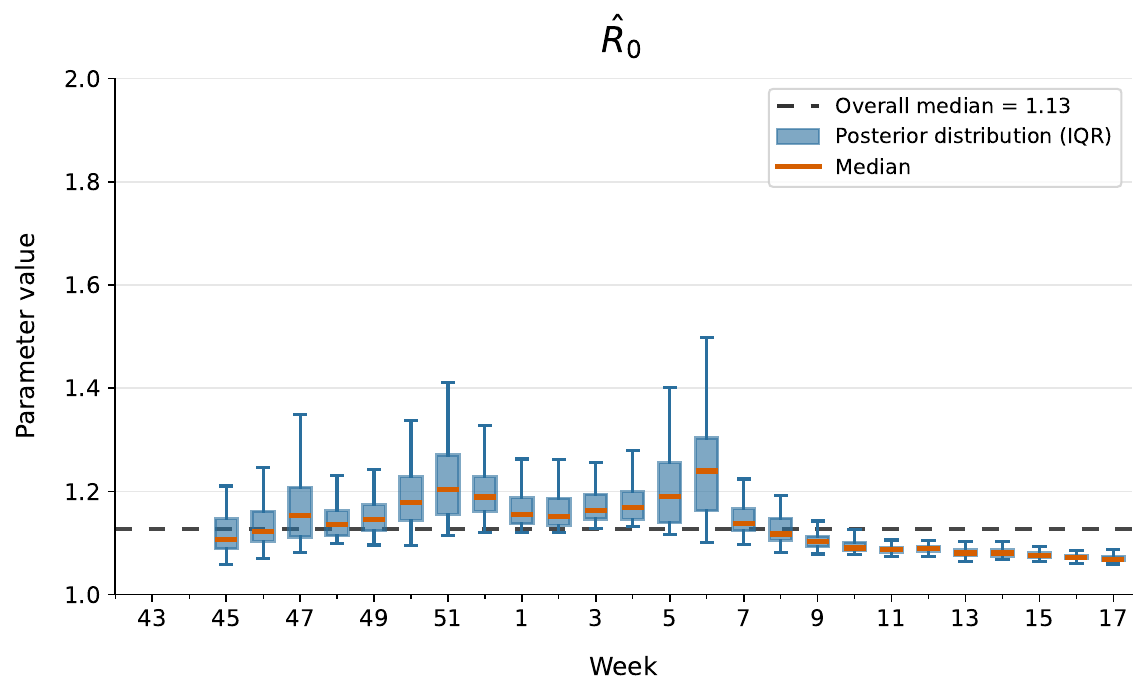}
    \quad
    \includegraphics[width=0.85\textwidth,keepaspectratio]{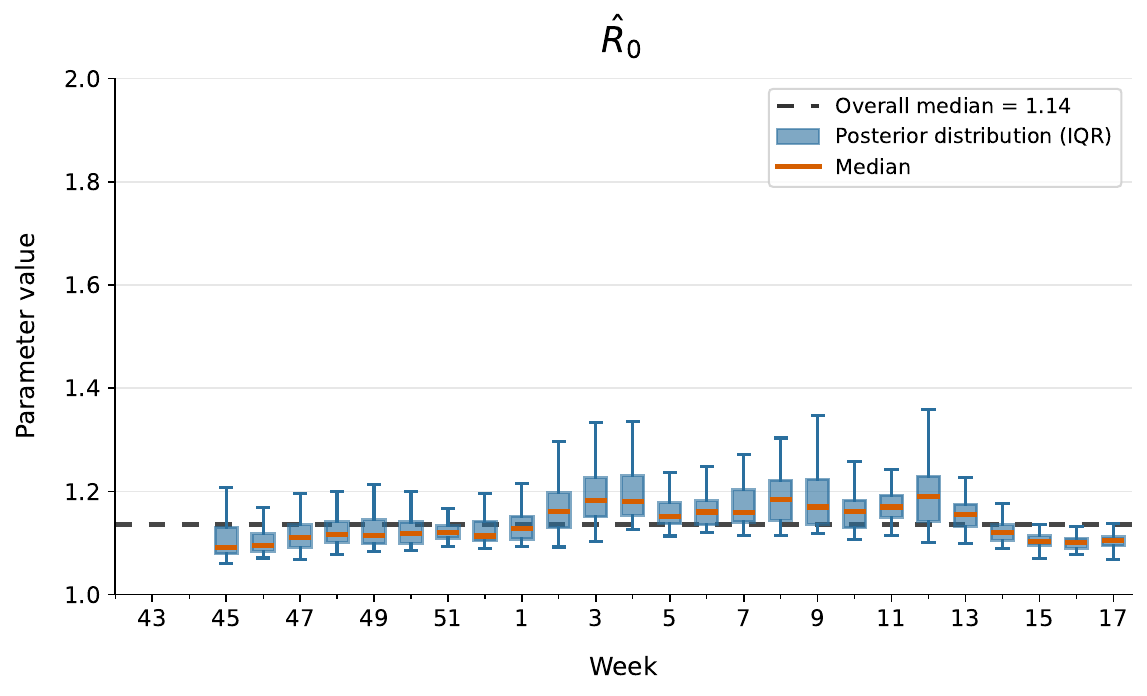}
\caption{\textbf{Behavior of the \gls{sir}-based model parameters $R_0$ via \gls{mcmc}.} Posterior distributions of the inferred parameter $\hat{R}_0 = \hat{\beta}/\hat{\gamma}$, basic reproduction number, across epidemic weeks. Each box represents the interquartile range (IQR, 25th–75th percentile), with whiskers extending to the 5th and 95th percentiles; the orange horizontal line within each box denotes the weekly posterior median. The dashed line in the panels indicates the overall posterior median across all weeks. Upper panel: national data from the 2023-2024 seasonal influenza. Lower panel: national data from the 2024-2025 seasonal influenza.}
\label{fig:Influcast_params_inference_all_R0}
\end{figure}

\begin{figure}[H]
\centering
\includegraphics[width=\textwidth,height=10cm]{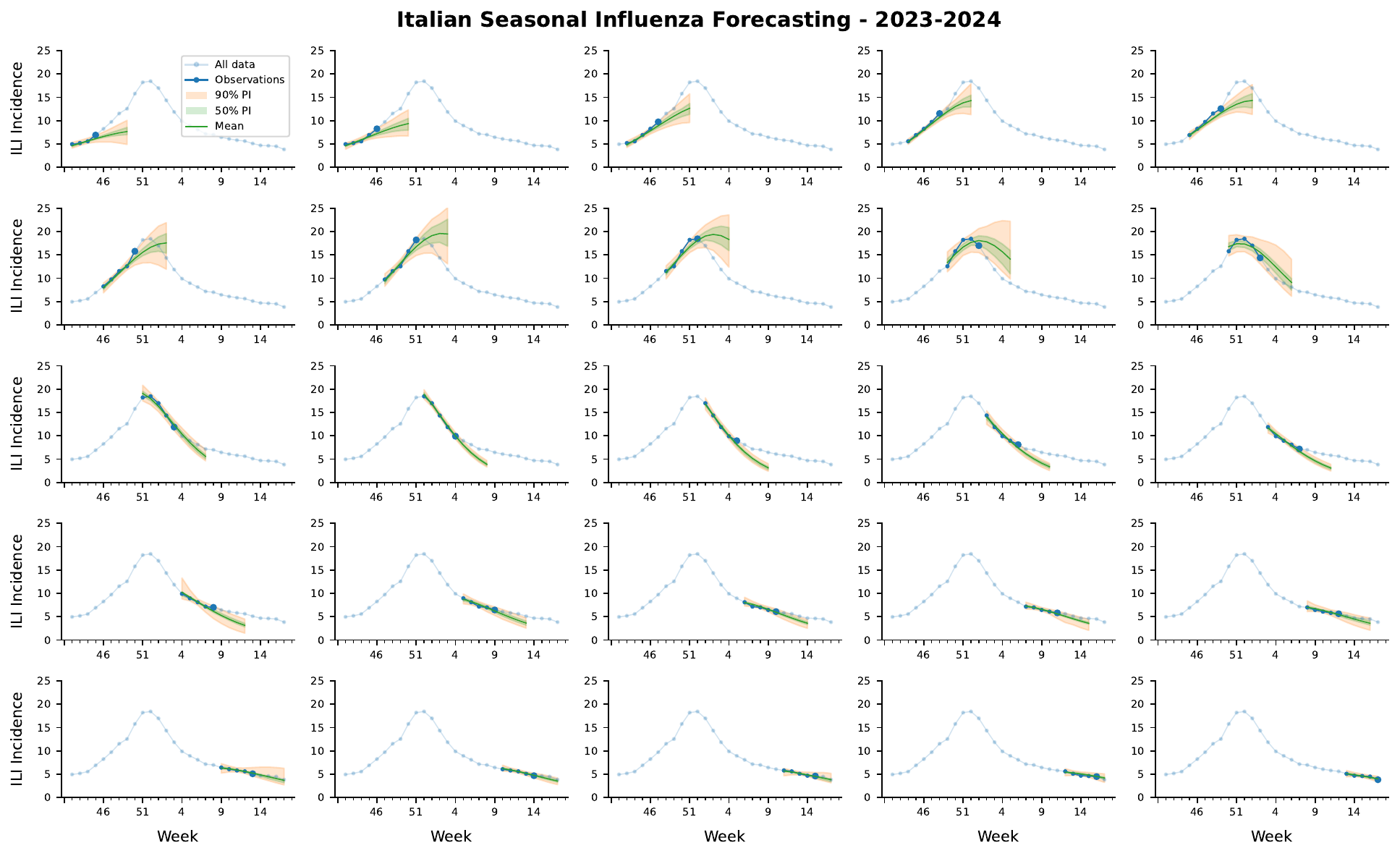}
\caption{\textbf{Seasonal influenza forecasting four weeks ahead via SIR-INN.} The ground truth data, in blue, represent the weekly observations at national level of the 2023-2024 seasonal influenza. The highlighted blue data represent the time window of observations, with fixed length of $5$, from which the \gls{sir}-based \gls{pinn} infers the epidemic parameters and then forecasts on the following four weeks ahead. The mean-model fits and forecasts are represented with green lines and the $50\%$ and $90\%$ percentile intervals are shaded, respectively, in green and orange error bars.}
\label{fig:Influcast_fit_all}
\end{figure}

\begin{figure}[H]
\centering
\includegraphics[width=\textwidth,height=10cm]{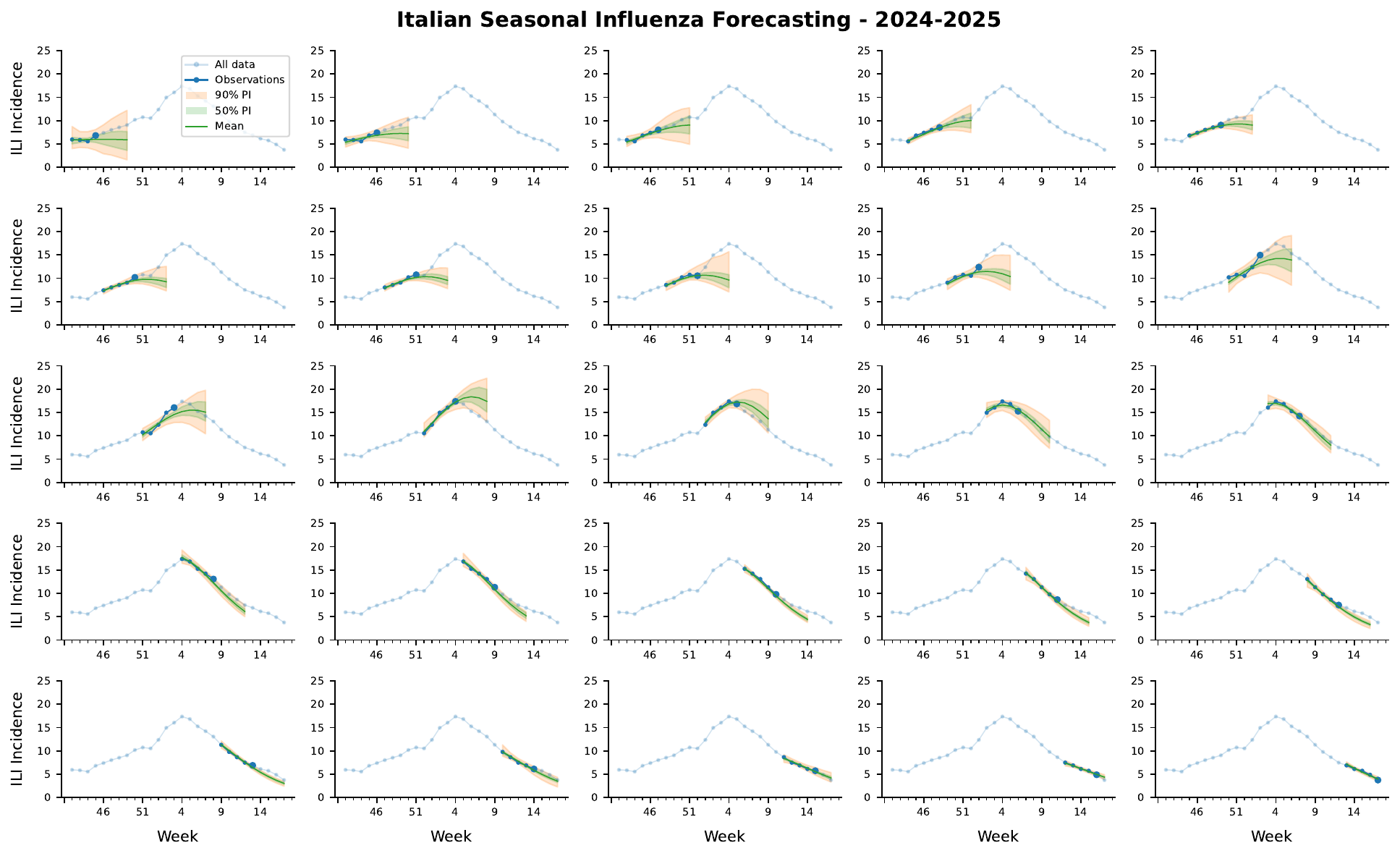}
\caption{\textbf{Seasonal influenza forecasting four weeks ahead via SIR-INN.} The ground truth data, in blue, represent the weekly observations at national level of the 2024-2025 seasonal influenza. The highlighted blue data represent the time window of observations, with fixed length of $5$, from which the \gls{sir}-based \gls{pinn} infers the epidemic parameters and then forecasts on the following four weeks ahead. The mean-model fits and forecasts are represented with green lines and the $50\%$ and $90\%$ percentile intervals are shaded, respectively, in green and orange error bars.}
\label{fig:Influcast_fit_all_25}
\end{figure}

\begin{figure}[H]
\centering
\includegraphics[width=\textwidth,keepaspectratio]{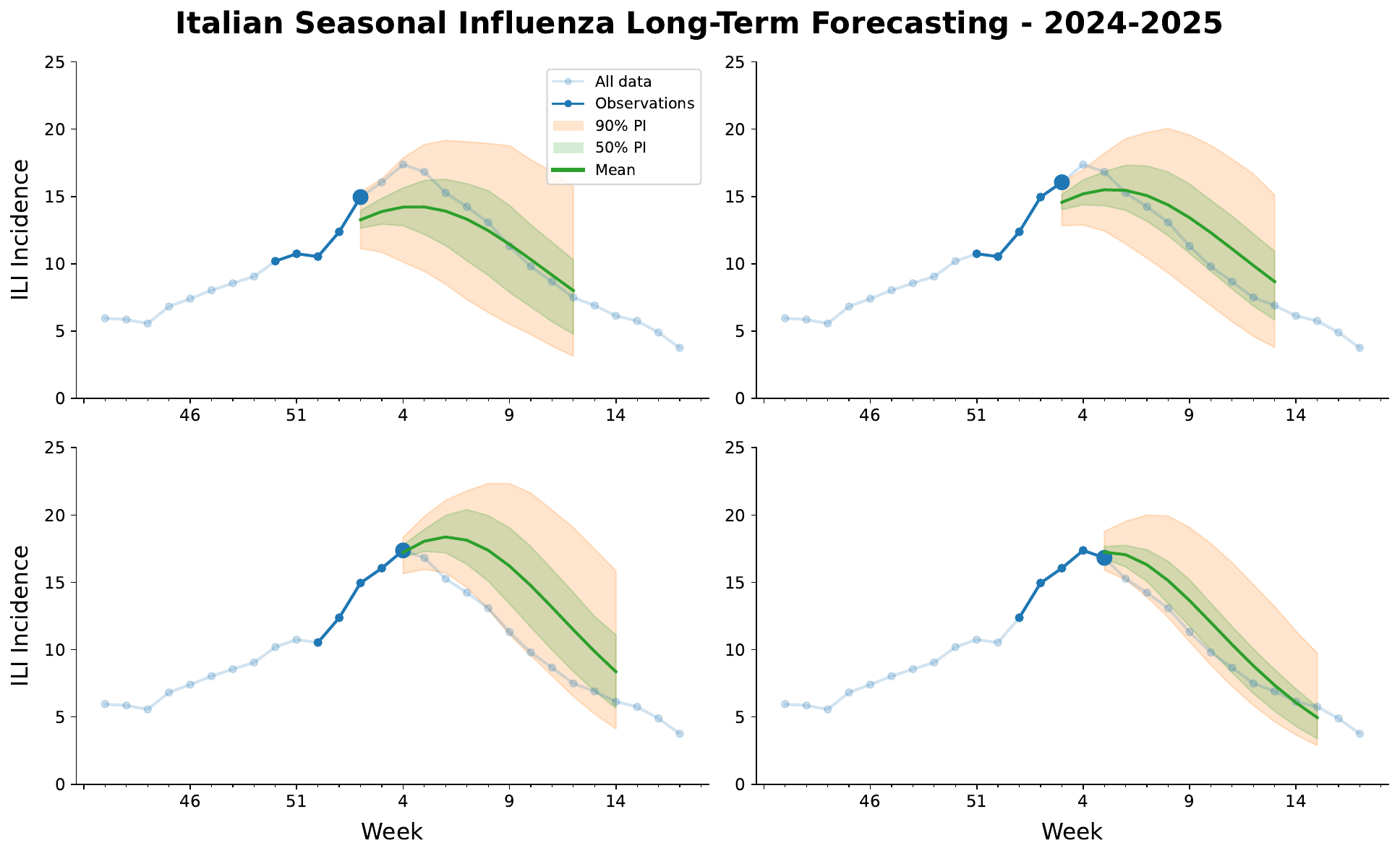}
\caption{\textbf{Seasonal influenza forecasting ten weeks ahead via SIR-INN.} The ground truth data, in blue, represent the weekly observations at national level of the 2024-2025 seasonal influenza. The highlighted blue data represent the time window of observations, with fixed length of $5$, from which the \gls{sir}-based \gls{pinn} infers the epidemic parameters and then forecasts on the following ten weeks ahead. The mean-model forecasts are represented with green lines and the $50\%$ and $90\%$ percentile intervals are shaded, respectively, in green and orange error bars.}
\label{fig:Influcast_long_forecasting_25}
\end{figure}

\clearpage
\newpage
\section{Population Conservation}\label{sec:N_const}

As mentioned in Section~\ref{sec: SIR-PINN}, a key structural property of the \gls{sir} model is population conservation which means $S(t)+I(t)+R(t)=1 \;\;\forall t \in [t_0,T]$ under the used population normalization (see also~\eqref{eq: model_SIR}).

Although this constraint was not explicitly enforced during training, we verify whether the trained neural network satisfies it.
For this, for each scenario in the training set we compute the absolute conservation error $|S(t)+I(t)+R(t)-1|$ at every training time point, and average these values to obtain a corresponding \gls{mae}, maximum absolute error, and standard deviation across time instants. 

Table~\ref{tab:conservation} reports these values.
The results show that SIR-INN recovers population conservation to a high degree of numerical accuracy across all tested epidemic scenarios, with a global \gls{mae} on the order of $10^{-4}$, suggesting that the conservation law emerges as an implicit property of the learned dynamics rather than requiring explicit enforcement.

\begin{table}[ht]
\centering
\small
\begin{tabular}{lccc}
\toprule
\textbf{Scenario} & \textbf{MAE} & \textbf{max $|$err$|$} & \textbf{std} \\
\midrule
\textbf{Global} & \num{2.3991e-04} & \num{2.0211e-02} & \num{4.3496e-04} \\
\bottomrule
\end{tabular}
\caption{Population conservation error globally computed over the full training set $\mathcal X_{\train}$ \eqref{eq: pinn_training_set}.}
\label{tab:conservation}
\end{table}

\clearpage
\newpage
\section{Generalization and out-of-distribution analysis}
\label{sec: B_ood}
We investigate in this section the suitability of the choice of the training set, especially regarding out-of-distribution behaviors of the network.

\begin{figure}[H]
\centering
\includegraphics[width=0.9\textwidth]{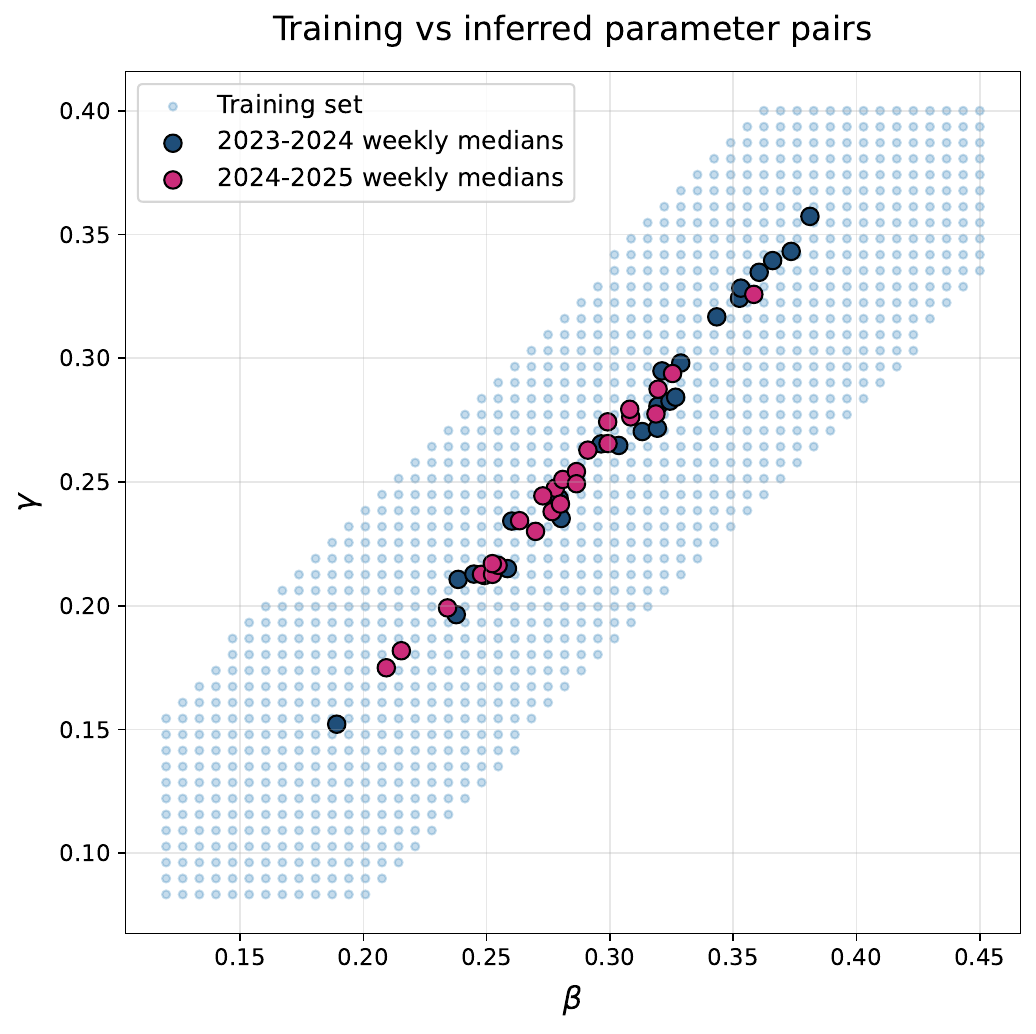}

\caption{\textbf{Training parameters and inferred parameters}. Grid of training parameter pairs (light-blue dots) and parameter pairs inferred by SIR-INN in each week for the 2023-2024 (blue dots) and 2024-2025 seasons (magenta dots).}
\label{fig:trainvsinfer}
\end{figure}

As detailed in Section~\ref{sec:train_details}, the SIR-PINN network is trained on pairs $(\beta, \gamma)$, whose grid location are reported as light-blued dots in Figure~\ref{fig:trainvsinfer}. These values are selected in accordance with the epidemiological literature, as discussed in Section~\ref{sec:train_details}.
The same figure shows also the location of the $(\hat{\beta},\hat{\gamma})$ parameters inferred by SIR-INN - specifically, the medians of the Markov chains estimated at each week of the influenza seasons 2023-2024 (blue dots) and 2024-2025 (magenta dots).
These estimated values clearly lie well within the boundaries of the training set, demonstrating that the choice of the ranges of the training parameters is appropriate.
We also stress that the inferred parameters in general do not coincide with the training locations, demonstrating the generalization ability of our framework.

\begin{figure}[H]
\centering
\includegraphics[
width=1.1\textwidth,
trim=3cm 4cm 3cm 3cm, 
clip
]{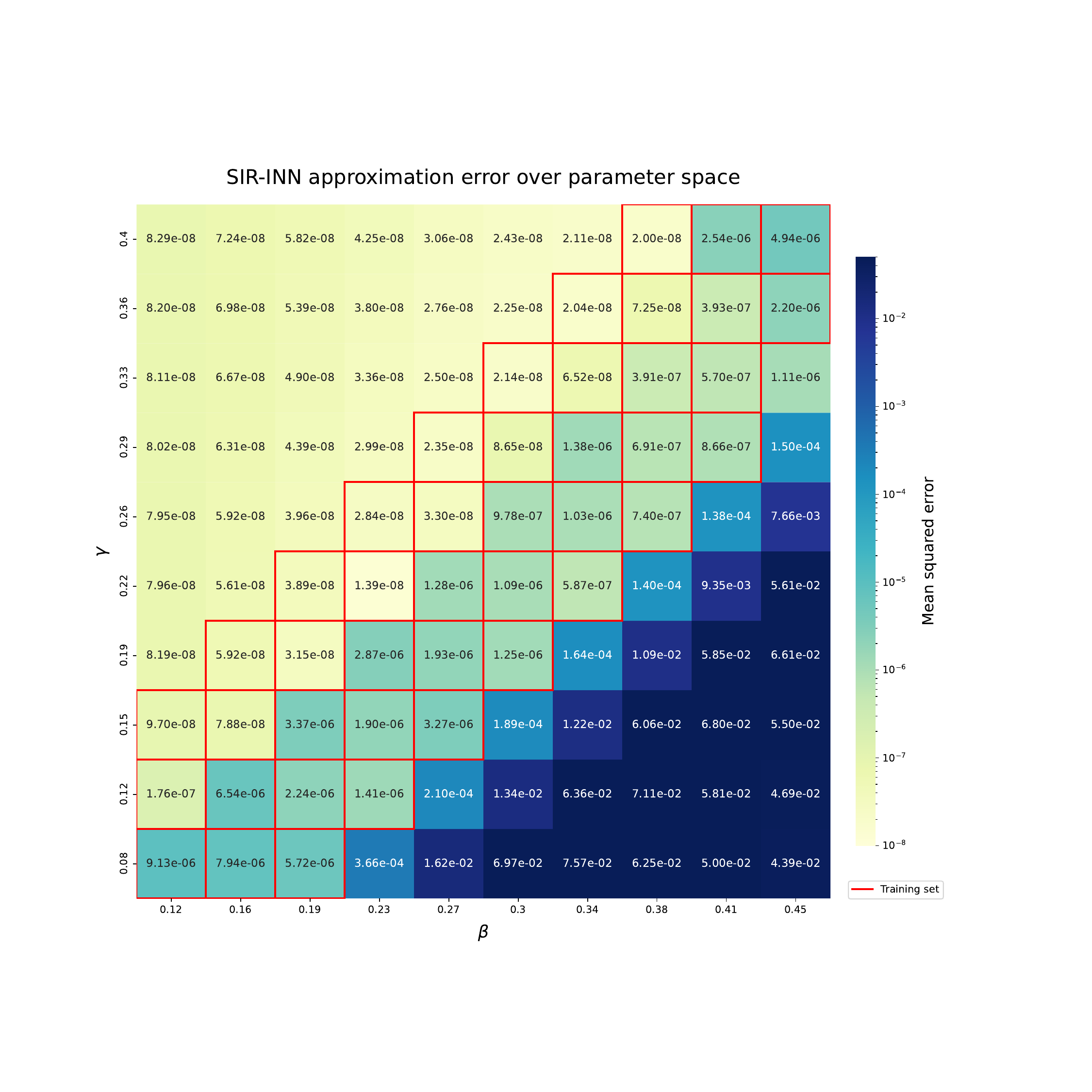}

\caption{\textbf{Approximation error for the SIR-INN framework}.
Error between the true \gls{sir} outputs and those predicted by the trained SIR-INN, computed on a regular grid of parameters $(\beta,\gamma)$. Red boxes indicate parameter pairs inside the training domain.}
\label{fig:error_heatmap}
\end{figure}

Figure~\ref{fig:error_heatmap} shows instead the behavior of the SIR-PINN in out of distribution regimes, namely, the \gls{mse} between the network prediction and the exact \gls{sir} values for a given parameter pair.
More precisely, we consider a regular grid of $10\times 10$ values $(\beta,\gamma)$ in the intervals  $B = [0.12, 0.45]$ and $\Gamma = [1/12,1/2.5]$.
The error measure is the \gls{mse}, normalized by $N$, computed over the entire time interval $[0,600]$, and averaged across the three compartments $S,I,R$.
The results are reported as a heatmap in Figure~\ref{fig:error_heatmap}, where each parameter pair is represented as a box containing the error value and colored according to the error magnitude. Boxes corresponding to parameters in the training area are framed in red.
Errors evaluated on parameter pairs that fall inside the training area are between $10^{-8}$ and $10^{-6}$, and extremely small errors are observed also in the out-of-distribution region with $R_0 = \beta/\gamma \ll 1$ (i.e., above the training diagonal). In this regime the epidemic does not occur and the \gls{sir} trajectories remain nearly constant over time, making them easier to approximate for the neural network.
Moving instead to the region with $R_0 = \beta/\gamma \gg 1$ (i.e., below the training diagonal), the approximation accuracy gradually decreases, as should be expected. Nevertheless, even in the most unfavorable cases, the error remains under values on the order of $10^{-2}$.

Taken together, these results show that the epidemiological parameters inferred from real influenza data for both seasons lie well within the training domain, a region where the neural approximation is accurate. This confirms the good choice of the training set that covers a wide region of the parameter space encompassing a broad range of epidemiological scenarios.

\clearpage
\newpage
\section{Robustness with respect to \gls{pinn} modeling assumptions}\label{sec:robust_modeling}
\subsection{Choice of the population size}\label{sec:population_size}
The \gls{sir} data used to train the \gls{pinn} (see Section~\ref{sec:train_details}) are constructed using a population size $N=10^6$. 
This choice is motivated by the fact that assuming the Italian population to be fully susceptible and well-mixed at the start of each influenza season is unrealistic.
Since we lack reliable estimates of pre-existing immunity and vaccination coverage, we assume $N$ to be an effective representation of the susceptible population. 
Furthermore, and crucially, our method is quite insensitive to the precise value for $N$. 
Indeed, the \gls{pinn} loss function was implemented with the \gls{sir} model governed by proportions (\ref{eq: model_SIR}), and also parameter inference is performed by re-scaling the measured data and the \gls{pinn} predictions (Section \ref{sec: detail_params}). As a consequence, both the learned dynamics, and the inferred parameters are not affected by the rescaling factor $N$.

To further support this point, we report in Table~\ref{tab: SIRINN 2seasons_large_population} the same analysis performed in Table~\ref{tab: SIRINN 2seasons}, but using a model trained with a larger population size ($N=10^7$).
The results show only minor differences across all metrics, confirming the robustness of the proposed method with respect to the choice of $N$.

\begin{table}[tbp]
\centering
\resizebox{\textwidth}{!}{
\begin{tabular}{lcccccc}
\toprule
\textbf{Model} & \textbf{Influenza Season} & \textbf{N. of Rounds} & \textbf{MAE} & \textbf{WIS} & \textbf{Coverage 50\%} & \textbf{Coverage 90\%} \\
\midrule
Baseline & 2023/2024 & 20 & 2.68 & 1.90 & \textbf{0.49} & \textbf{0.66} \\
SIR-INN & 2023/2024 & 20 & 2.28 & 1.63 & 0.11 & 0.39 \\
SIR-INN: $N=10^7$ & 2023/2024 & 20 & \textbf{2.07} & \textbf{1.40} & 0.15 & \textbf{0.66} \\
\midrule
Baseline & 2024/2025 & 21 & 2.54 & 1.71 & 0.14 & 0.52 \\
SIR-INN & 2024/2025 & 21 & \textbf{1.98} & \textbf{1.36} & \textbf{0.24} & 0.58 \\
SIR-INN: $N=10^7$ & 2024/2025 & 21 & 2.05 & 1.68 & 0.20 & \textbf{0.63} \\
\bottomrule
\end{tabular}
}
\caption{\textbf{Forecasting performance of the SIR-INN model, compared to its modified version (i.e., $N=10^7$), and to the baseline.} The results are obtained from the four-weeks-ahead forecasting at national level, across the two influenza seasons, in terms of \gls{mae} of the median forecast, \gls{wis}, $50\%$ and $90\%$ coverage. Bold values indicate the best performance for each metric within the corresponding season.}
\label{tab: SIRINN 2seasons_large_population}
\end{table}

\subsection{Effect of stochastic perturbations in \gls{pinn} training}\label{sec:stochastic_training}
As explained in Section \ref{sec:train_details}, the \gls{pinn} model proposed in this work is trained on deterministic synthetic data generated from the \gls{sir} system, with the goal of accurately approximating the underlying ODE dynamics. In our framework, the stochasticity of real-world epidemiological observations is not modeled at the training stage, but is instead explicitly accounted for during parameters inference step. We derived incidence from the susceptible trajectory through a finite-difference approximation, i.e., it is proportional to the variation of $S$ over time. As a consequence, injecting independent noise at the compartment level would propagate through this operation and lead to amplified fluctuations in the resulting incidence, potentially introducing uncontrolled variability rather than improving robustness. Furthermore, adding count-based noise (e.g., Poisson noise or stochastic simulations via Gillespie's algorithm) would therefore introduce a mismatch between the training data and the model assumptions, and may also violate structural constraints such as conservation of the total population.

For completeness, we conducted an additional experiment in which the synthetic training trajectories were perturbed with bounded stochastic noise applied to the normalized $S$, $I$, and $R$ components. Specifically, for each time point, we sample independent uniform noise in the interval $[-\epsilon, \epsilon]$ (with $\epsilon = 0.02$ in our experiments) and scale it proportionally to the magnitude of each normalized compartment $S$, $I$, and $R$. To preserve the population conservation constraint, we enforce a zero-sum correction by subtracting the mean of the perturbations across the three components, then adjusting the perturbed compartments to ensure positivity, and finally renormalizing them so that their sum remains equal to one. The perturbation level was chosen to mimic realistic reporting variability while preserving epidemiological consistency.

The corresponding quantitative results on four-weeks ahead Italian seasonal influenza forecasting are shown in Table~\ref{tab: SIRINN 2seasons_noisy_data}. We report the same analysis performed in Table~\ref{tab: SIRINN 2seasons}, but using a model trained with perturbed synthetic data.
Overall, the performance of the model remains comparable to that obtained with deterministic training data, with no significant improvements observed across all evaluation metrics.
Given the absence of measurable benefits and the additional computational cost associated with stochastic perturbations, we opted for the deterministic training strategy presented in the main experiments.

\begin{table}[tbp]
\centering
\resizebox{\textwidth}{!}{
\begin{tabular}{lcccccc}
\toprule
\textbf{Model} & \textbf{Influenza Season} & \textbf{N. of Rounds} & \textbf{MAE} & \textbf{WIS} & \textbf{Coverage 50\%} & \textbf{Coverage 90\%} \\
\midrule
Baseline & 2023/2024 & 20 & 2.68 & 1.90 & \textbf{0.49} & \textbf{0.66} \\
SIR-INN & 2023/2024 & 20 & 2.28 & 1.63 & 0.11 & 0.39 \\
SIR-INN: Noisy data & 2023/2024 & 20 & \textbf{2.13} & \textbf{1.56} & 0.14 & 0.45 \\
\midrule
Baseline & 2024/2025 & 21 & 2.54 & 1.71 & 0.14 & 0.52 \\
SIR-INN & 2024/2025 & 21 & \textbf{1.98} & \textbf{1.36} & \textbf{0.24} & 0.58 \\
SIR-INN: Noisy data & 2024/2025 & 21 & 2.16 & 1.54 & 0.21 & \textbf{0.64} \\
\bottomrule
\end{tabular}
}
\caption{\textbf{Forecasting performance of the SIR-INN model, compared to its modified version (i.e., Noisy data), and to the baseline.} The results are obtained from the four-weeks-ahead forecasting at national level, across the two influenza seasons, in terms of \gls{mae} of the median forecast, \gls{wis}, $50\%$ and $90\%$ coverage. Bold values indicate the best performance for each metric within the corresponding season.}
\label{tab: SIRINN 2seasons_noisy_data}
\end{table}

\subsection{Hyperparameter tuning investigation}\label{sec:param_losses}
As discussed in Section~\ref{sec: SIR-PINN}, the PINN loss~\eqref{eq: Loss_PINN_general} is sometimes defined by considering a different weighting of the two loss components, i.e.,
\begin{equation*}
\mathcal{L} = \mathcal{L}_{data} + \alpha_{ODE}\cdot\mathcal{L}_{ODE},
\end{equation*}
where $\alpha_{ODE}> 0$ is a positive weight balancing the effect of the two components.

Although the results in the main text are obtained with $\alpha_{ODE}=1=10^0$ (uniform weighting of the two components), we test values
$\alpha_{ODE} = 10^{-2}, 10^{-1}, 10^1, 10^{2}$.

To evaluate the performances of this hyperparameter choice, we retrain a new \gls{pinn} for each value of $\alpha_{ODE}$ following the training process and all remaining settings described in Section \ref{sec: SIR-PINN} and Section~\ref{sec:train_details}. Then we evaluate the corresponding error in approximating the exact \gls{sir} components for a set of parameter pairs. More specifically, the pairs are those defined in~\ref{sec: B_ood}, which fall inside the training area (please also refer to Figure \ref{fig:error_heatmap}), and approximation errors are computed as described in the same~\ref{sec: B_ood}.

\begin{figure}[htbp]
    \centering

    \begin{subfigure}[t]{0.49\textwidth}
        \centering
        \includegraphics[width=\textwidth]{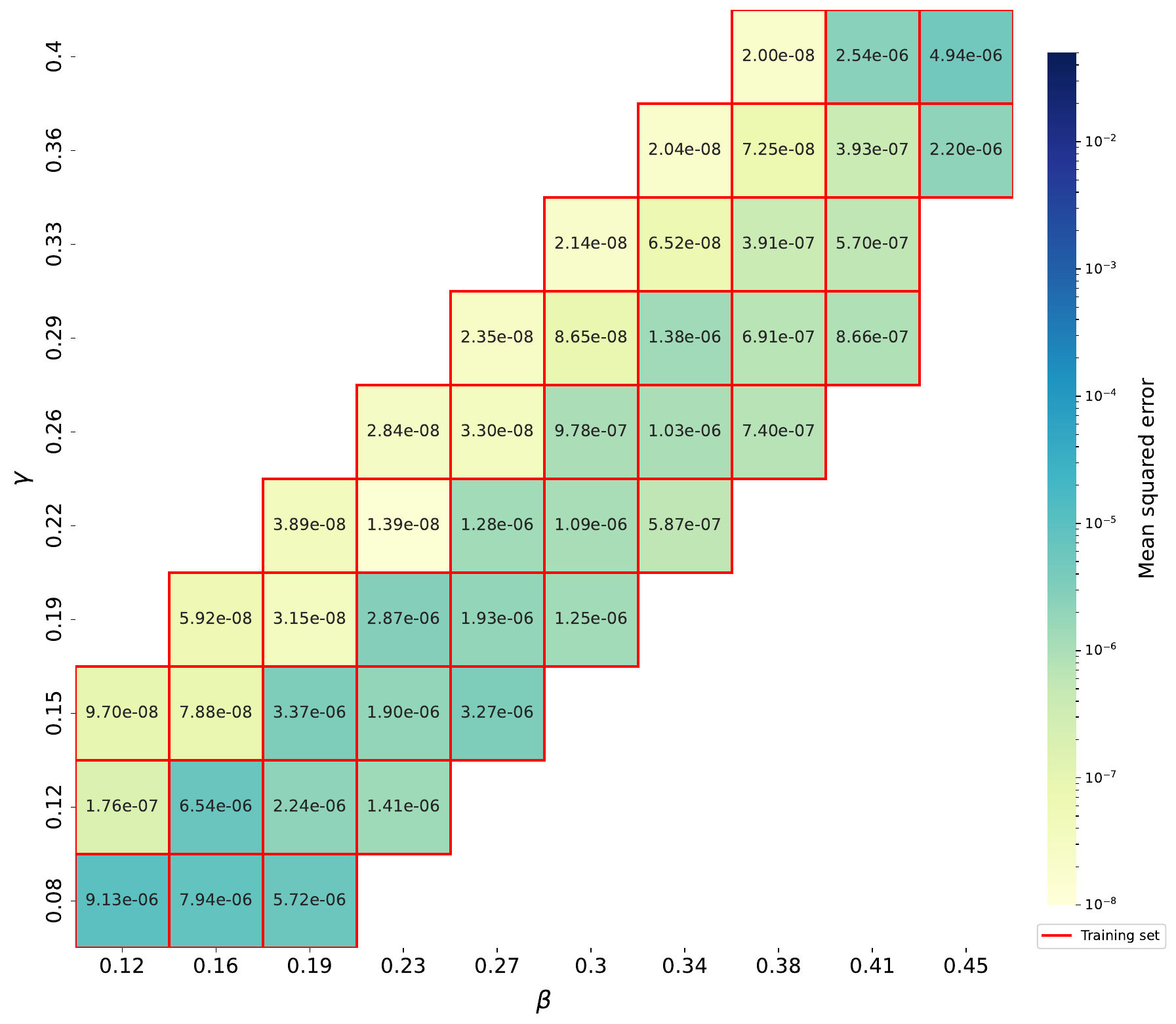}
        \caption{PINN baseline: $\alpha_{ODE} = 1$}
        \label{fig:pinn_base}
    \end{subfigure}

    \vspace{0.8em}

    \begin{subfigure}[t]{0.49\textwidth}
        \centering
        \includegraphics[width=\textwidth]{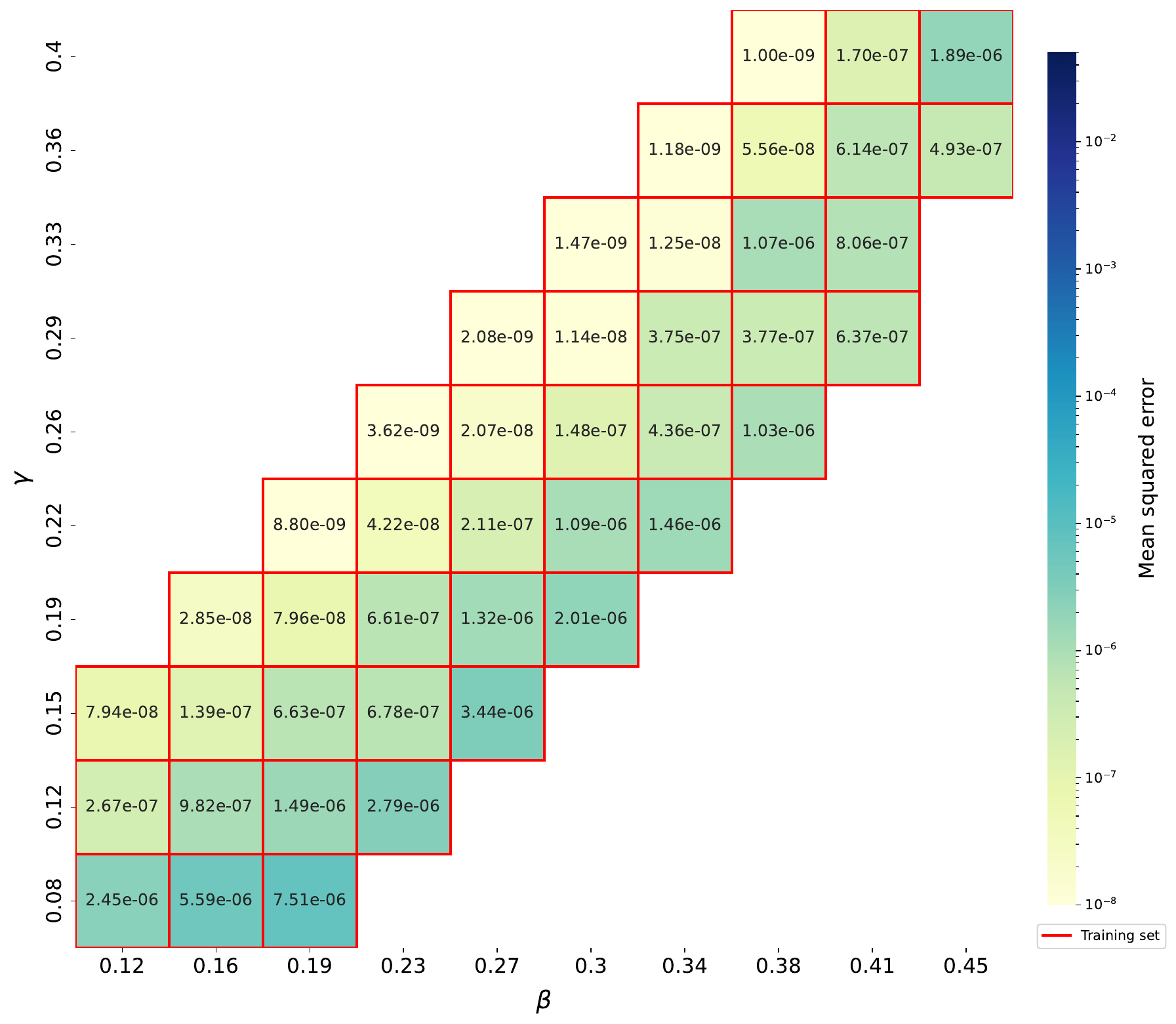}
        \caption{PINN: $\alpha_{ODE} = 0.01$}
        \label{fig:pinn_1}
    \end{subfigure}
    \hfill
    \begin{subfigure}[t]{0.49\textwidth}
        \centering
        \includegraphics[width=\textwidth]{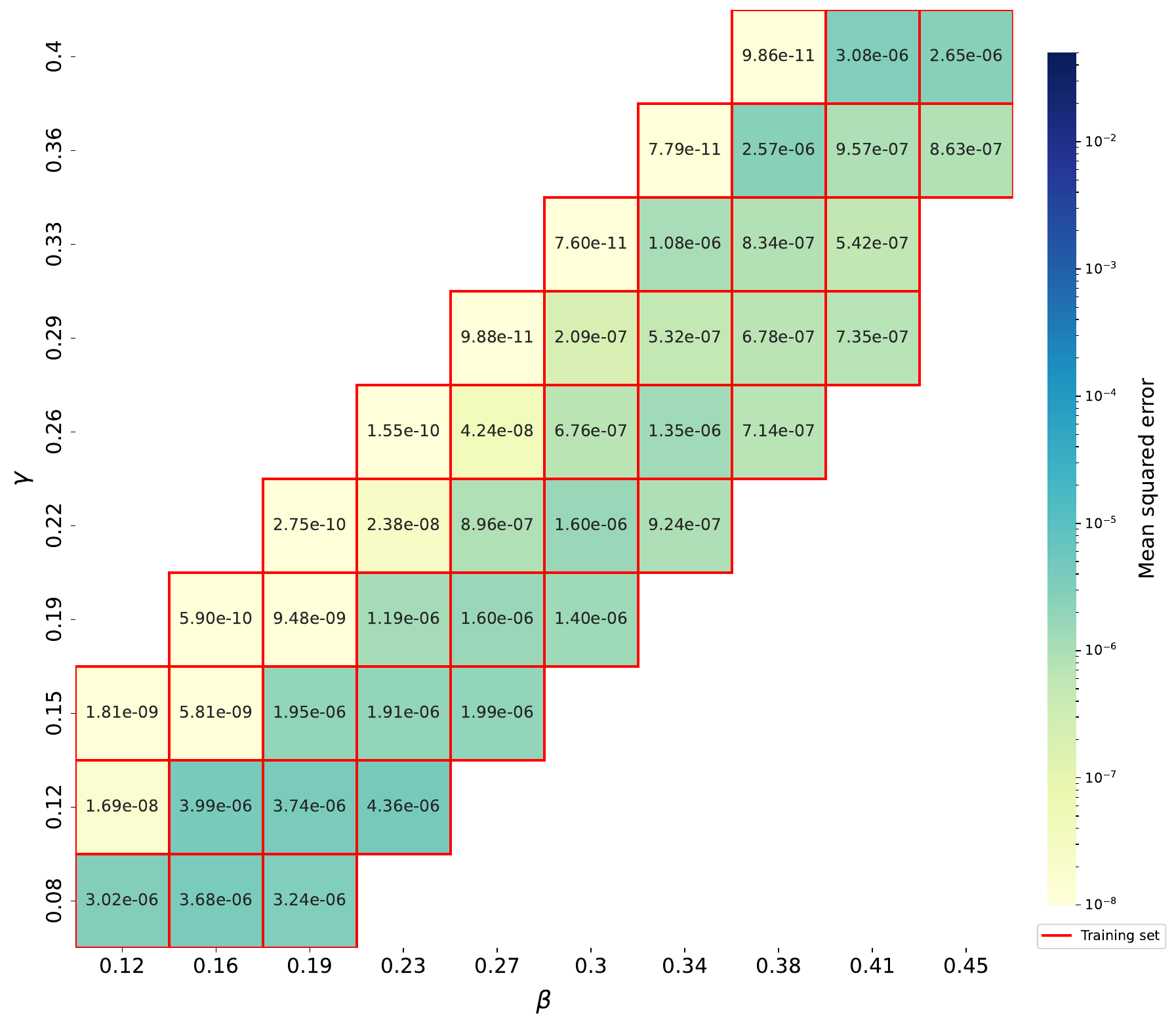}
        \caption{PINN: $\alpha_{ODE} = 0.1$}
        \label{fig:pinn_2}
    \end{subfigure}

    \vspace{0.4em}

    \begin{subfigure}[t]{0.49\textwidth}
        \centering
        \includegraphics[width=\textwidth]{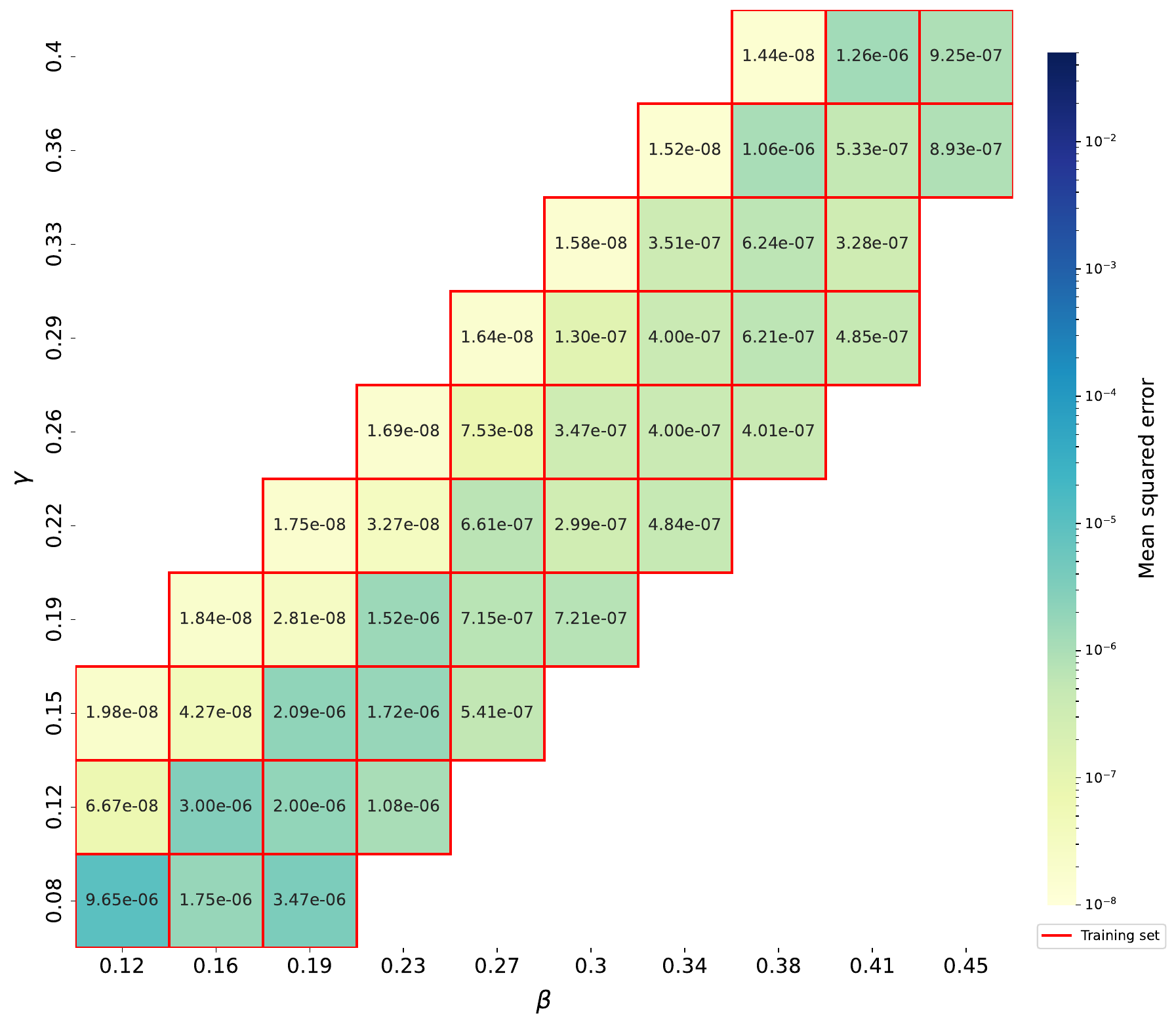}
        \caption{PINN: $\alpha_{ODE} = 10$}
        \label{fig:pinn_3}
    \end{subfigure}
    \hfill
    \begin{subfigure}[t]{0.49\textwidth}
        \centering
        \includegraphics[width=\textwidth]{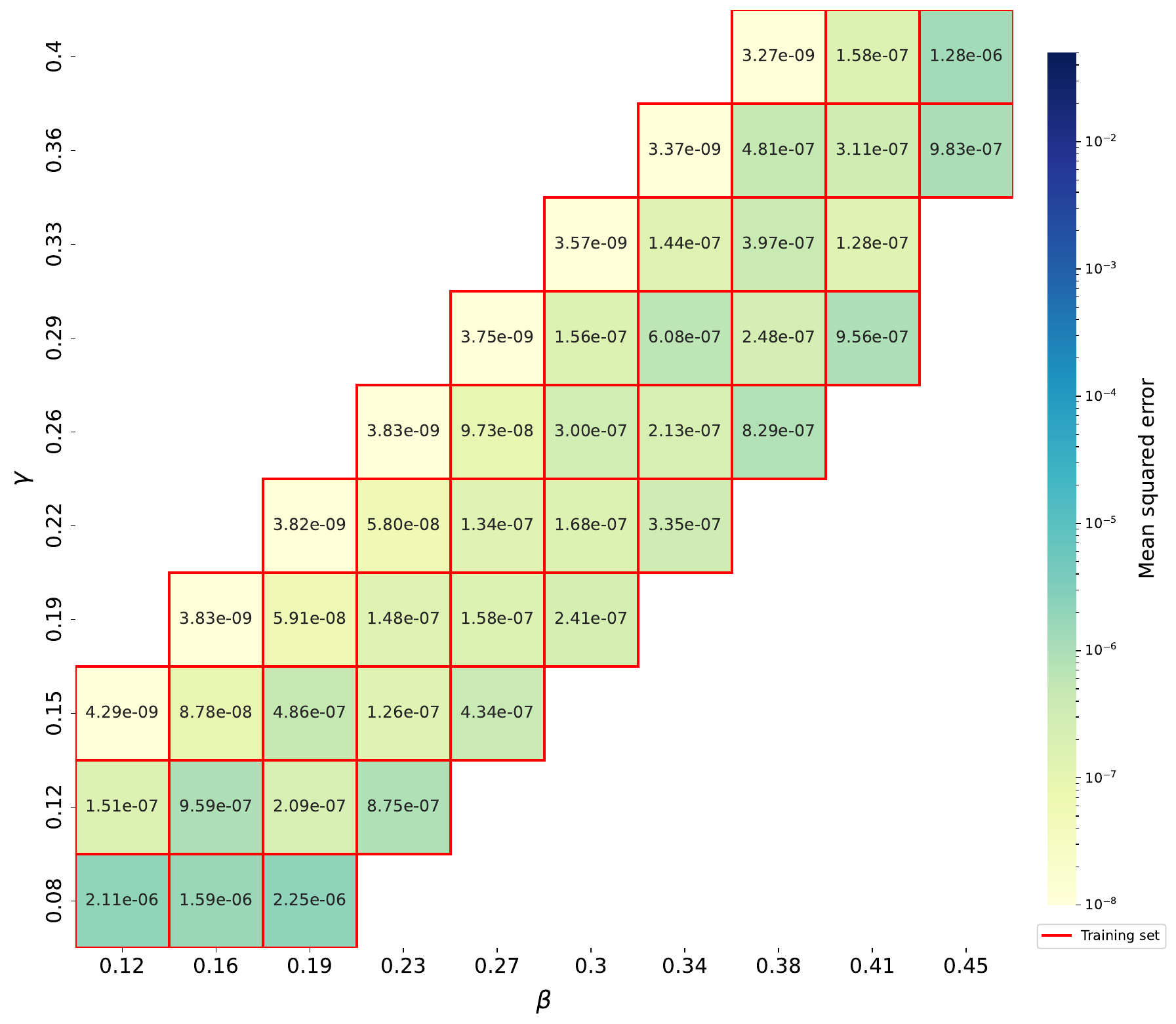}
        \caption{PINN: $\alpha_{ODE} = 100$}
        \label{fig:pinn_4}
    \end{subfigure}

    \caption{Comparison of approximation error distributions across PINNs over parameter pairs inside the training domain. Top: PINN baseline model used as reference. Bottom: four architecture variants.}
    \label{fig:pinn_comparison}

\end{figure}

\begin{figure}[tp]
\centering
    \centering
    \includegraphics[width=\textwidth,keepaspectratio]{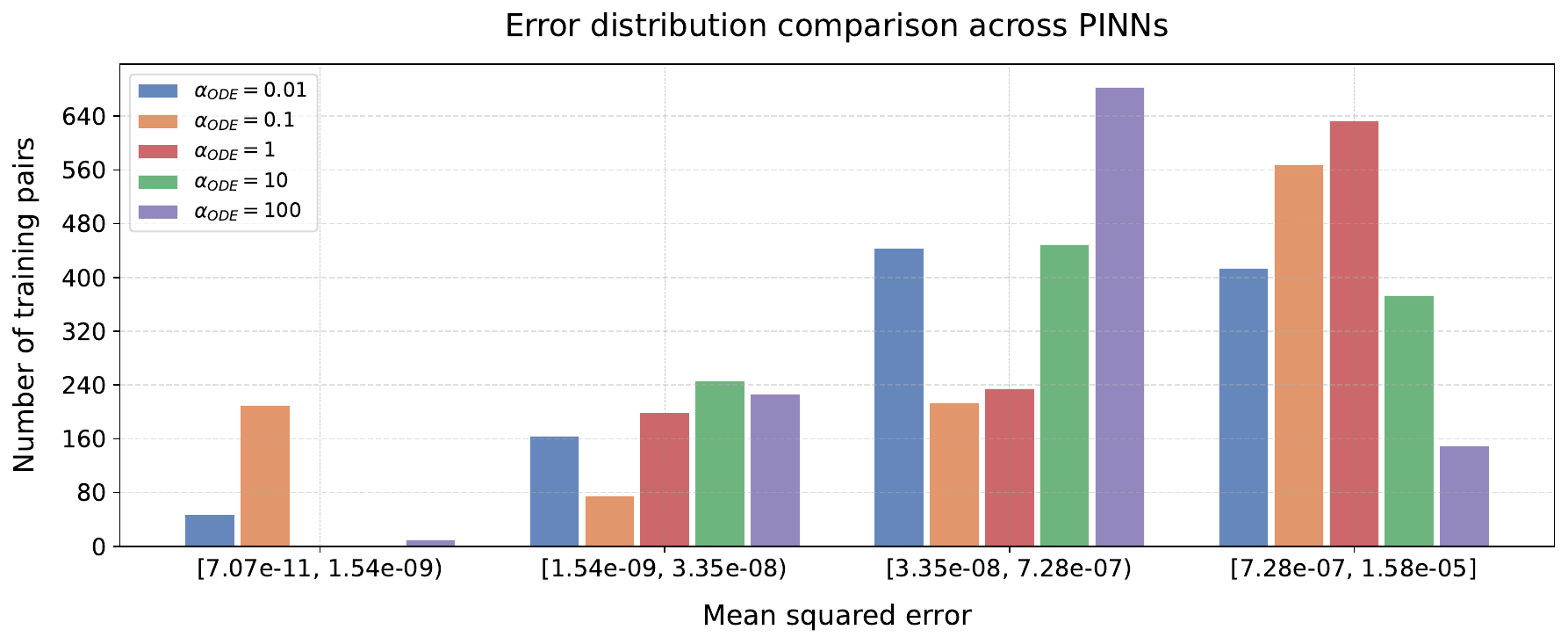}
\caption{Distribution of the mean squared error (averaged over $S$, $I$, and $R$ components) across all training parameter pairs $(\beta, \gamma)$, for five PINN configurations differing in hyperparameter settings (i.e., weights of the ODE component of the loss function). Each bin reports the number of pairs falling in the corresponding error range.}
\label{fig:bar_plot_losses}
\end{figure}

Figure~\ref{fig:pinn_comparison} reports these results for $\alpha_{ODE}=1$ (our choice) and for the other values of $\alpha_{ODE}$. It appears clear that, although limited differences are visible, the overall error performance is essentially analogous for any choice of $\alpha_{ODE}$. To further validate this observation, Figure~\ref{fig:bar_plot_losses} shows the distribution of the errors computed for all the training pairs $(\beta,\gamma)$, highlighting once more that there is a significant overlap of the error distributions for different values of $\alpha_{ODE}$.

Since no parameter provides a significant advantage, we prefer to choose $\alpha_{ODE}=1$, which provides a more interpretable and clear definition of the loss~\eqref{eq: Loss_PINN_general}.

\clearpage
\newpage
\section{Robustness of the \gls{mcmc} parameter estimation}
\subsection{Impact of chain length}\label{sec:chain_length}
To validate the choice of the length of the Markov Chain ($10,000$, see Section~\ref{sec: detail_params}) we repeat all inference experiments using instead $50,000$ iterations. For simplicity, we restrict to the 2023-2024 season, although analogous results are observed for the 2024-2025 season.

\begin{figure}[h!]
\centering
\hspace*{-0.1\linewidth}
\includegraphics[width=1.15\textwidth,height=10cm]
{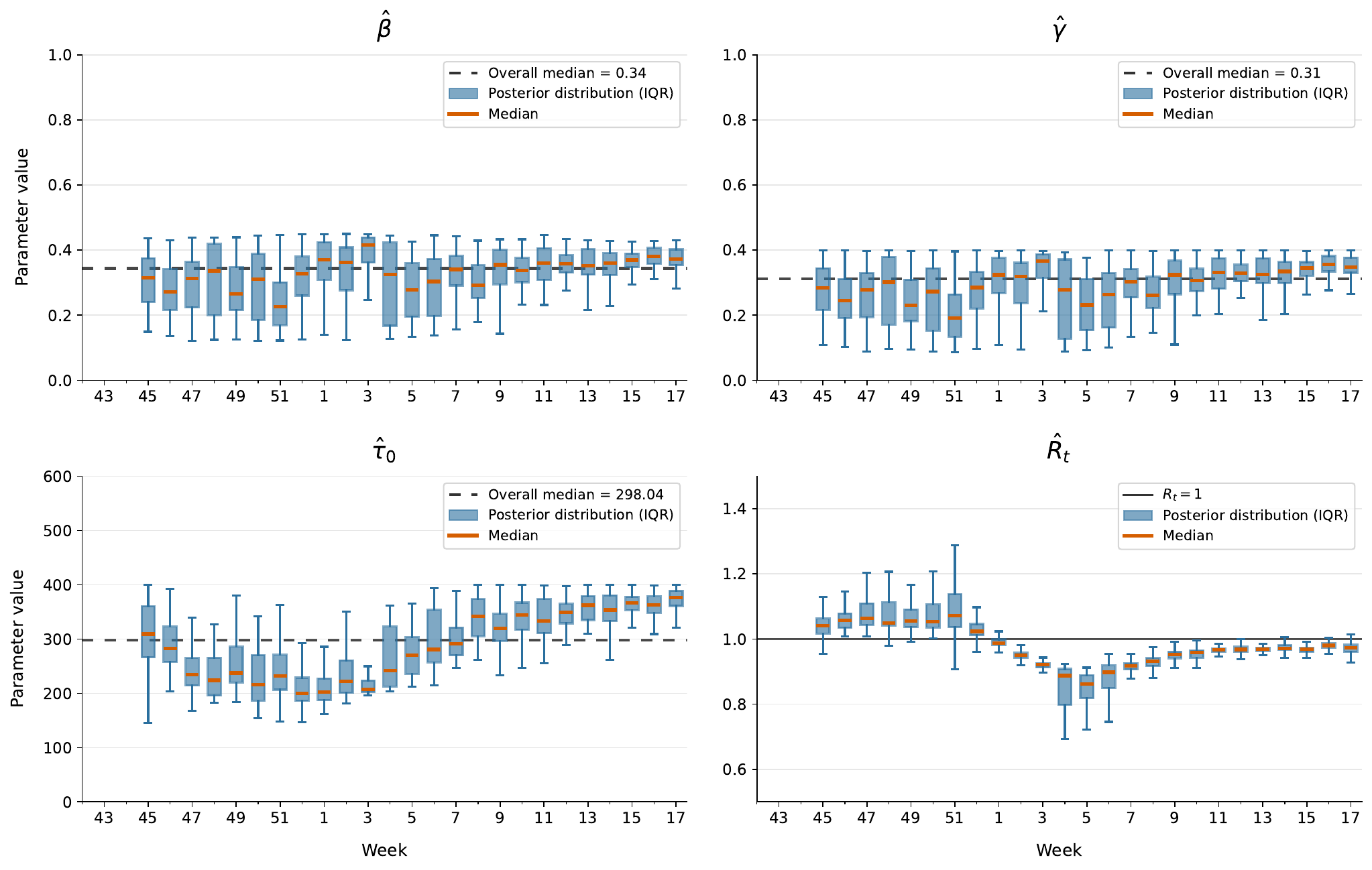}
\caption{\textbf{Behavior of the SIR-based model parameters estimated via MCMC.} At each week we consider the observations time window of size 5 which includes the previous four weeks and the current one. The inferred features represented in the figure are, respectively, the transmission rate, the recovery rate, the initial time of the spreading, and the effective reproduction number. The results correspond to an augmented chain length of $50,000$, compared to $10,000$ used in Figure~\ref{fig:Influcast_params_inference}. National data used are from seasonal influenza of 2023-2024.}
\label{fig:Influcast_params_inference_all_50000}
\end{figure}

\begin{figure}[h!]
\centering
    \centering
    \includegraphics[width=\textwidth,keepaspectratio]{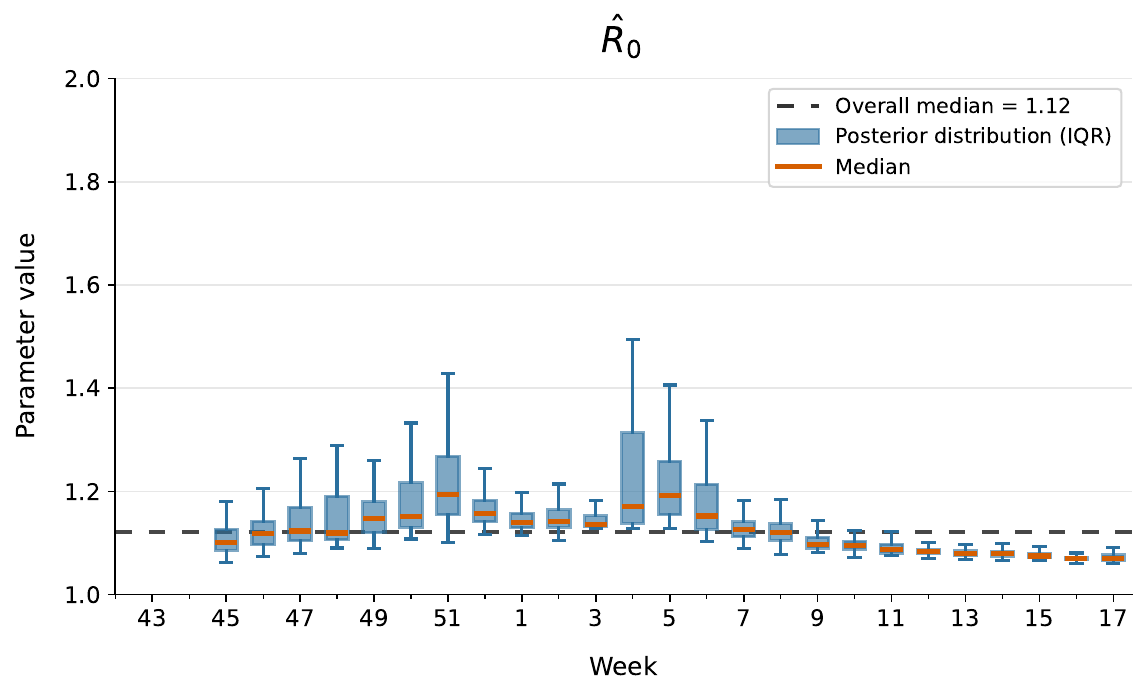}
\caption{\textbf{Behavior of the SIR-based model parameters R0 via MCMC.}
At each week we consider the observations time window of size 5 which includes the previous four weeks and the current one. The results correspond to an augmented chain length of $50,000$, compared to $10,000$ used in Figure~\ref{fig:Influcast_params_inference_all_R0} (upper panel). National data used are from seasonal influenza of 2023-2024.}
\label{fig:Influcast_params_inference_all_R0_50000}
\end{figure}

The corresponding inference results are reported in Figure~\ref{fig:Influcast_params_inference_all_50000} and Figure~\ref{fig:Influcast_params_inference_all_R0_50000}. Comparing these estimates with Figure~\ref{fig:Influcast_params_inference} and Figure~\ref{fig:Influcast_params_inference_all_R0} (upper panel), we see that no noticeable difference can be detected, concluding that a length of $10,000$ is sufficient to reach convergence of \gls{mcmc}. 
Given that shorter chains significantly reduce computational cost, we adopt this configuration value in all experiments presented in the paper.

\subsection{Impact of the prior choice}\label{sec:Gaussian priors}

In the baseline configuration, we assume uniform priors for all model parameters. While weakly informative priors are often recommended in modern Bayesian practice, this choice is motivated by the specific structure of our inference framework.

As explained in detail in Section \ref{sec: detail_params}, our approach relies on a rolling-window estimation procedure, where the SIR parameters ($\beta$, $\gamma$, $\tau_0$) are re-estimated at each week using only the most recent observations (a window of size $M=5$). This design reflects the non-stationary nature of influenza dynamics, where transmission and recovery patterns may vary significantly over time due to behavioral changes, interventions, or viral evolution. As a result, we effectively fit independent SIR models at each forecasting round, rather than a single model with time-varying parameters.
As a consequence, incorporating prior information could artificially constrain the parameter space and potentially introduce bias when the underlying dynamics change. For this reason, we adopt uniform priors to allow the data within each observation window to drive the inference with minimal prior assumptions.

To assess the robustness of the inference with respect to the prior specification, we perform the same analysis reported in Section \ref{sec:4weeks} using weakly informative Gaussian priors. Specifically, we modify the prior distributions for the parameters $\beta$ and $\gamma$ only, replacing the uniform priors with Gaussian ones. The means are set to the midpoint of the intervals used as constraints in the \gls{mcmc} implementation (i.e., the initial guesses), while the standard deviations are chosen as one third of the corresponding interval widths. For the temporal parameter $\tau_0$, we leave the uniform prior assumption, as there is no motivation to support a more structured distribution.

We report in Table~\ref{tab: SIRINN 2seasons_Gaussian_priors} the same analysis performed in Table~\ref{tab: SIRINN 2seasons}, but using Gaussian priors instead of uniform priors for the epidemiological parameters.

The results show only minor differences in forecasting performance, indicating that the Poisson likelihood driven by the observed data dominates the inference process. These findings confirm that the proposed framework is robust with respect to the choice of prior. \\
While informative priors may be beneficial in settings where reliable prior knowledge is available or when modeling a full epidemic season jointly, uniform priors provide a flexible and appropriate choice for our rolling-window forecasting strategy.

\begin{table}[tbp]
\centering
\resizebox{\textwidth}{!}{
\begin{tabular}{lcccccc}
\toprule
\textbf{Model} & \textbf{Influenza Season} & \textbf{N. of Rounds} & \textbf{MAE} & \textbf{WIS} & \textbf{Coverage 50\%} & \textbf{Coverage 90\%} \\
\midrule
Baseline & 2023/2024 & 20 & 2.68 & 1.90 & \textbf{0.49} & \textbf{0.66} \\
SIR-INN & 2023/2024 & 20 & 2.28 & \textbf{1.63} & 0.11 & 0.39 \\
SIR-INN: Gaussian priors & 2023/2024 & 20 & \textbf{2.21} & 1.70 & 0.11 & 0.43 \\
\midrule
Baseline & 2024/2025 & 21 & 2.54 & 1.71 & 0.14 & 0.52 \\
SIR-INN & 2024/2025 & 21 & \textbf{1.98} & \textbf{1.36} & \textbf{0.24} & \textbf{0.58} \\
SIR-INN: Gaussian priors & 2024/2025 & 21 & 2.02 & 1.54 & 0.20 & 0.54 \\
\bottomrule
\end{tabular}
}
\caption{\textbf{Forecasting performance of the SIR-INN model, compared to its modified version (i.e., Gaussian priors), and to the baseline.} The results are obtained from the four-weeks-ahead forecasting at national level, across the two influenza seasons, in terms of \gls{mae} of the median forecast, \gls{wis}, $50\%$ and $90\%$ coverage. Bold values indicate the best performance for each metric within the corresponding season.}
\label{tab: SIRINN 2seasons_Gaussian_priors}
\end{table}

\clearpage
\newpage
\section{Comparison with the mechanistic model}\label{sec:SIRvsODE}

This supplementary section provides a direct comparison between three \gls{sir}-based approaches for the four-weeks ahead forecasting of Italian seasonal influenza. Specifically, we compare the proposed PINN-based approach (our SIR-INN), a classical mechanistic SIR model solved via numerical integration with \texttt{odeint} (named ODE$\rightarrow$ODE), as well as a hybrid variant (named PINN$\rightarrow$ODE) in which PINN-inferred parameters are used to initialize and propagate the ODE forward. Notice that the \gls{sir} model is endowed with the same knowledge, constraints, and setting of the SIR-INN framework during the parameter inference step via MCMC, ensuring a fair and controlled comparison.

\begin{figure}[tp]
\centering
    \begin{subfigure}[b]{\textwidth} 
    \hspace*{-0.1\linewidth}
    \centering
    \includegraphics[width=1.2\textwidth,keepaspectratio]{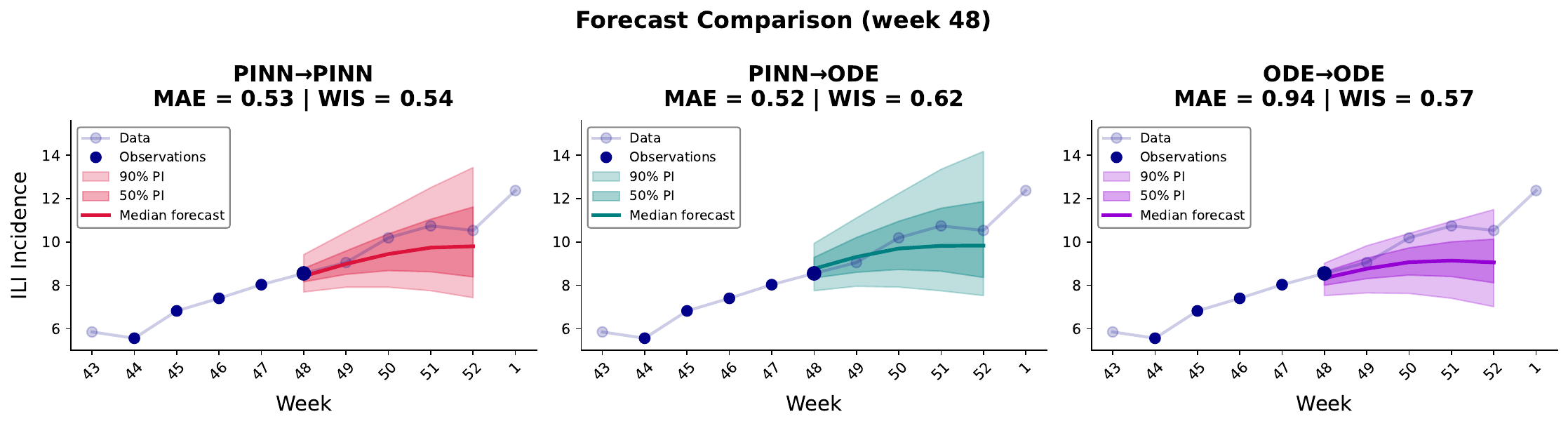}
    \end{subfigure}
    
\quad

    \begin{subfigure}[b]{\textwidth} 
    \hspace*{-0.1\linewidth}
    \centering
    \includegraphics[width=1.2\textwidth,keepaspectratio]{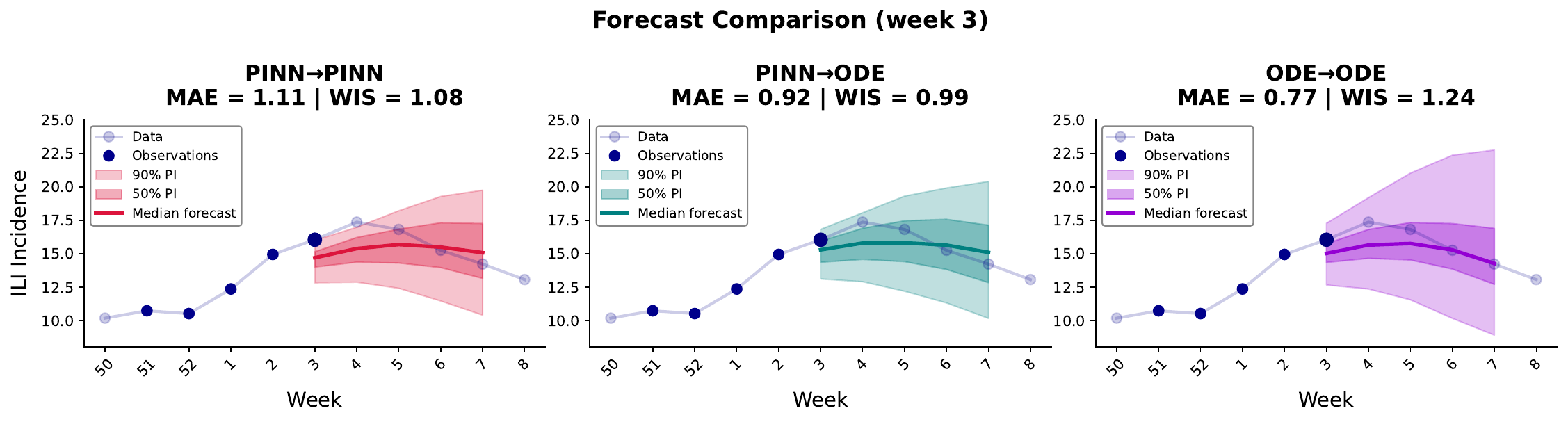}
    \end{subfigure}
    
\quad

    \begin{subfigure}[b]{\textwidth} 
    \hspace*{-0.1\linewidth}
    \centering
    \includegraphics[width=1.2\textwidth,keepaspectratio]{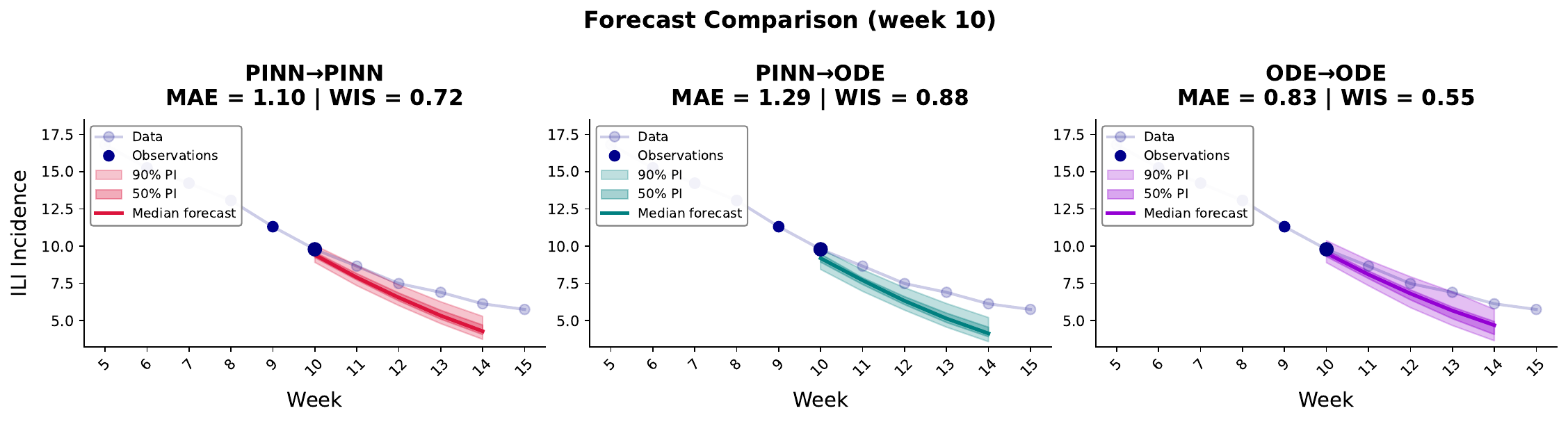}
    \end{subfigure}
\caption{\textbf{Four-weeks ahead probabilistic forecast comparison of 2024-2025 Italian seasonal influenza.} The three methodologies involved are: PINN$\rightarrow$PINN (first column), PINN$\rightarrow$ODE (second column), and ODE$\rightarrow$ODE (third column). The represented scenarios describe three different phases of the influenza: initial phase (upper panel), peak phase (middle panel), and decreasing stage (lower panel).}
\label{fig:SIRvsODE_plot}
\end{figure}

Table~\ref{tab: SIRvsODE_table} summarises the forecasting performance of the three approaches on the 2024--2025 Italian seasonal influenza season across the four metrics: \gls{mae}, \gls{wis}, and coverages at the 50\% and 90\% nominal levels. Figure~\ref{fig:SIRvsODE_plot} illustrates 
representative four-week-ahead probabilistic forecasts at three characteristic phases of the epidemic curve: the initial growth phase (upper panel), the epidemic peak (middle panel), and the declining stage (lower panel).\\
Overall, the three approaches yield comparable forecasting accuracy. The PINN$\rightarrow$PINN pipeline achieves the lowest \gls{mae} (1.98) and  \gls{wis} (1.36) along with the 
shortest total runtime (249.10\,s on par with PINN$\rightarrow$ODE), confirming its computational advantage over the classical ODE solver. This advantage is expected to become even more pronounced as model complexity increases and as the knowledge constraint decreases. The ODE$\rightarrow$ODE approach provides the highest 90\% coverage (0.64), but requires a substantially larger runtime (444.25,s). Finally, the hybrid PINN$\rightarrow$ODE approach achieves the same 50\% coverage (0.24) as PINN$\rightarrow$PINN, but ranks last in terms of \gls{wis}. 

The qualitative comparison in Figure~\ref{fig:SIRvsODE_plot} further highlights that differences across methods are most apparent during the 
initial phase (upper panel), where data uncertainty is highest, and tend to diminish closer to and after the peak (middle and lower panels). 

These results suggest that the choice of forecasting approach may be non-trivial and is unlikely to reduce to a single universally optimal strategy. Beyond the epidemiological context, this question touches on a broader methodological challenge inherent to hybrid frameworks that combine physics-informed neural networks with probabilistic inference. The interplay between the expressiveness of the surrogate model, the consistency with the mechanistic prior, and the uncertainty quantification introduced by the MCMC sampling step affects predictive performance in ways that are not immediately transparent. The relative merits of a fully neural, a fully mechanistic, or a hybrid pipeline may therefore shift depending on the specific epidemic phase, the forecast horizon, the target evaluation metric, and potentially changes across seasons, viruses, and regions. We recognize this as a delicate and multifaceted question that deserves a dedicated and systematic investigation, going well beyond the scope of the present methodological study. A thorough analysis of these trade-offs is currently underway and will be the subject of a follow-up work.

\begin{table}[ht]
\centering
\resizebox{\textwidth}{!}{
\begin{tabular}{lccccc}
\toprule
\textbf{Approach} & \textbf{MAE} & \textbf{WIS} & \textbf{Coverage 50\%} & \textbf{Coverage 90\%} & \textbf{Total Runtime (s)} \\
\midrule
PINN$\rightarrow$PINN & \textbf{1.98} & \textbf{1.36} & \textbf{0.24} & 0.58 & \textbf{249.10} \\
PINN$\rightarrow$ODE  & 2.03 & 1.40 & \textbf{0.24} & 0.56 & \textbf{249.10} \\
ODE$\rightarrow$ODE   & 2.00 & 1.49 & 0.13 & \textbf{0.64} & 444.25 \\
\bottomrule
\end{tabular}
}
\caption{Comparison of forecasting approaches for 2024-2025 Italian seasonal influenza. Best values for each metric are highlighted in bold.}
\label{tab: SIRvsODE_table}
\end{table}

\clearpage
\newpage
\section{Definition of the error metrics}
\label{sec:error_metrics_defs}
We report here a precise definition of the error metrics defined in Section~\ref{sec:metrics}. These details are taken from Section 4.3 in~\cite{fiandrino2025collaborative}, to which we refer for further discussion.

To introduce these definitions, we denote as $y$ the reference value (e.g., the measured incidence), and as $f$ a probabilistic forecast having median $m$.
The \gls{wis} is depending on a parameter $\alpha\in[0,1)$. Let $[\ell,u]$ with $\ell, u\in\R$ be the prediction interval of the forecast $f$ corresponding to $(1-\alpha)\cdot 100\%$. We first define the Interval Score as
\begin{equation*}
IS_{\alpha}(f, y)
=
\begin{cases}
u-\ell + \frac2\alpha(\ell-y), & y< \ell,\\
u-\ell, &\ell\leq y\leq u,\\
u-\ell + \frac2\alpha(y-u),&u<y.
\end{cases}
\end{equation*}
Since $\ell, u$ depend on the forecast, we explicitly indicate the dependency on $f$ in the left-hand side.
Then, the \gls{wis} is averaging this value over $K\in\N$ choices $\alpha_1, \dots, \alpha_K>0$. Letting $w_0, w_1, \dots, w_k>0$ be positive weights, we set
\begin{equation*}
WIS_{\alpha_1, \dots, \alpha_K}(f, y) = \frac{1}{K+\frac12} \cdot\left(w_0\cdot|y-m|+\sum_{k=1}^K \left(w_k\cdot IS_{\alpha_k}(f,y))\right)\right).
\end{equation*}
Following~\cite{fiandrino2025collaborative} we use
\begin{align*}
K &= 11,\\
\alpha_k&\in\{0.02, 0.05, 0.10, 0.20, 0.30, 0.40, 0.50, 0.60, 0.70, 0.80, 0.90\},\\
w_0 &= 1/2,\\
w_k &= \alpha_k/2.
\end{align*}

The \gls{ae} is instead computed as
\begin{equation*}
AE(f, y) = |m-y|.
\end{equation*}
Assuming a temporal sequence $y=(y_1, \dots, y_M)$ of observations, and corresponding forecasts $f=(f_1, \dots, f_M)$ with medians $m_1, \dots, m_M$, the \gls{mae} is simply averaging the \glspl{ae} as
\begin{equation*}
MAE(f, y) = \frac{1}{M}AE(f_i, y_i).
\end{equation*}

Finally, the coverage $C_\alpha(f, y)$ of level $\alpha\in [0,1)$ counts the number of times the observations $y_i$ fall inside the a prediction interval, averaged over the times $i=1, \dots, M$.  More precisely, considering again the interval $[\ell,u]$ corresponding to $(1-\alpha)\cdot 100\%$, we define
\begin{equation*}
C_\alpha(f, y)=  \frac{1}{M}\left|\left\{i\in\{1, \dots, M\}: l_i\leq y_i\leq u_i\right\}\right|.  
\end{equation*}
We consider the $50\%$ ($\alpha=0.5$) and $90\%$ ($\alpha=0.1$) coverages.





\end{document}